\documentclass[%
 reprint,
 amsmath,amssymb,
 aps,
]{revtex4-2}

\usepackage{graphicx}
\usepackage{dcolumn}
\usepackage{bm}
\usepackage{upgreek}
\usepackage{xcolor}
\usepackage{enumitem}
\usepackage{placeins}

\begin{document}

\title{Microscopic treatment of instantaneous spectral diffusion and its effect on quantum gate fidelities in rare-earth-ion-doped crystals}

\author{Adam Kinos}
\email{adam.kinos@fysik.lth.se}
\author{Lars Rippe}
\author{Andreas Walther}
\author{Stefan Kr\"{o}ll}

\affiliation{%
 Department of Physics, Lund University, P.O. Box 118, SE-22100 Lund, Sweden
}%

\date{\today}

\begin{abstract}
The effect of instantaneous spectral diffusion (ISD) on gate operations in rare-earth-ion-doped crystals is an important question to answer for the future of rare-earth quantum computing. Here we present a microscopic modeling that highlights the stochastic nature of the phenomenon, and use it to investigate ISD errors on single-qubit gate operations. Furthermore, we present a method to estimate the total error from many different error sources by only studying subsystems containing one error source at a time. This allows us to estimate the total ISD error from all non-qubit dopants in the vicinity of a qubit. We conclude that optical pumping techniques must be used to empty the frequency regions around the qubit transitions from absorption (transmission windows) in order to suppress the ISD errors. Despite using such windows, there remains a roughly $0.3\%$ risk that a qubit has an ISD error larger than the error from other sources. In those cases, the qubit can be discarded and its frequency channel can be reused by another qubit. However, in most cases the ISD errors are significantly smaller than other errors, thus opening up the possibility to perform noisy intermediate-scale quantum (NISQ) algorithms despite ISD being present. 
\end{abstract}

\maketitle

\section{\label{sec:intro}Introduction}
Rare-earth-ion-doped crystals are versatile materials that have been used in, e.g., quantum memories \cite{Nilsson2005a, Kraus2006, Hetet2008, Riedmatten2008, Afzelius2010, Hedges2010, Beavan2012a, Sabooni2013a, Dajczgewand2014, Guendogan2015, Jobez2015, Schraft2016}, conversion between optical and microwave signals \cite{OBrien2014, Williamson2014, FernandezGonzalvo2019}, and quantum computing \cite{Pryde2000, Ichimura2001, Ohlsson2002, Nilsson2002, Wesenberg2003, Wesenberg2007, Walther2009a, Walther2015, Ahlefeldt2020, Grimm2021, Hizhnyakov2021, Kinos2021a}. This has in large part been thanks to their long life- \cite{Konz2003} and coherence times \cite{Equall1994, Sun2002, Zhong2015}, and their capacity to store large amounts of information. 

In these systems, spectral diffusion, which causes shifts to the transition frequencies of ions, can often be an unwanted effect. In this work, we study a form of spectral diffusion called instantaneous spectral diffusion (ISD), where an incident light field alters the excitation of an ion which in turn shifts the optical transitions of other nearby ions \cite{Mims1961, Klauder1962, Mims1968, Huang1989, Huang1990, Equall1994}. A good overview of the history of ISD for rare-earth-ions can be found in, e.g., Ref. \cite{Thiel2014a}. Since most experiments so far have been performed on ensembles of rare-earth-ions, the theoretical investigations of ISD mostly describe the effect on average \cite{Klauder1962, Liu1990, Liu1990a, Thiel2014a}, where it is observed as a dephasing mechanism that depends on the degree of excitation. However, for applications that rely on single ions, e.g., the future of rare-earth quantum computing \cite{Wesenberg2007, Kinos2021}, ISD must be analyzed on the microscopic scale of the individual ions. 

This work consists of two main parts. First, Secs. \ref{sec:general_ISD} and \ref{sec:extrapolation_method} show how one can simulate ISD, how ISD affects a gate operation in an idealized system, and present the \textit{\textbf{Q}ubit \textbf{B}loch vector estimation based on \textbf{I}ndependent \textbf{E}rror \textbf{S}ources} (QBies) method, which is used to estimate the total error from many different error sources. This part is more general, especially the QBies method which can be used to investigate any system where errors from different sources are mostly independent.

The second part, presented in Secs. \ref{sec:ISD_Crystal} and \ref{sec:ISD_crystal_analysis}, use the QBies method to study ISD in one possible implementation of a rare-earth quantum computer. Specifically, we investigate how ISD affects a single-qubit (SQ) NOT operation using the pulses designed in Ref. \cite{Kinos2021a}. Our intention is that the work presented in this part can act as a foundation on how to study the effects of ISD on a single-ion level even when considering more complicated gate operations. Finally, to set the stage for this part we end the introduction by providing a brief overview of the relevant parts of the envisioned rare-earth quantum computer, for more information see Ref. \cite{Kinos2021}. 

\begin{figure*}
\includegraphics[width=\textwidth]{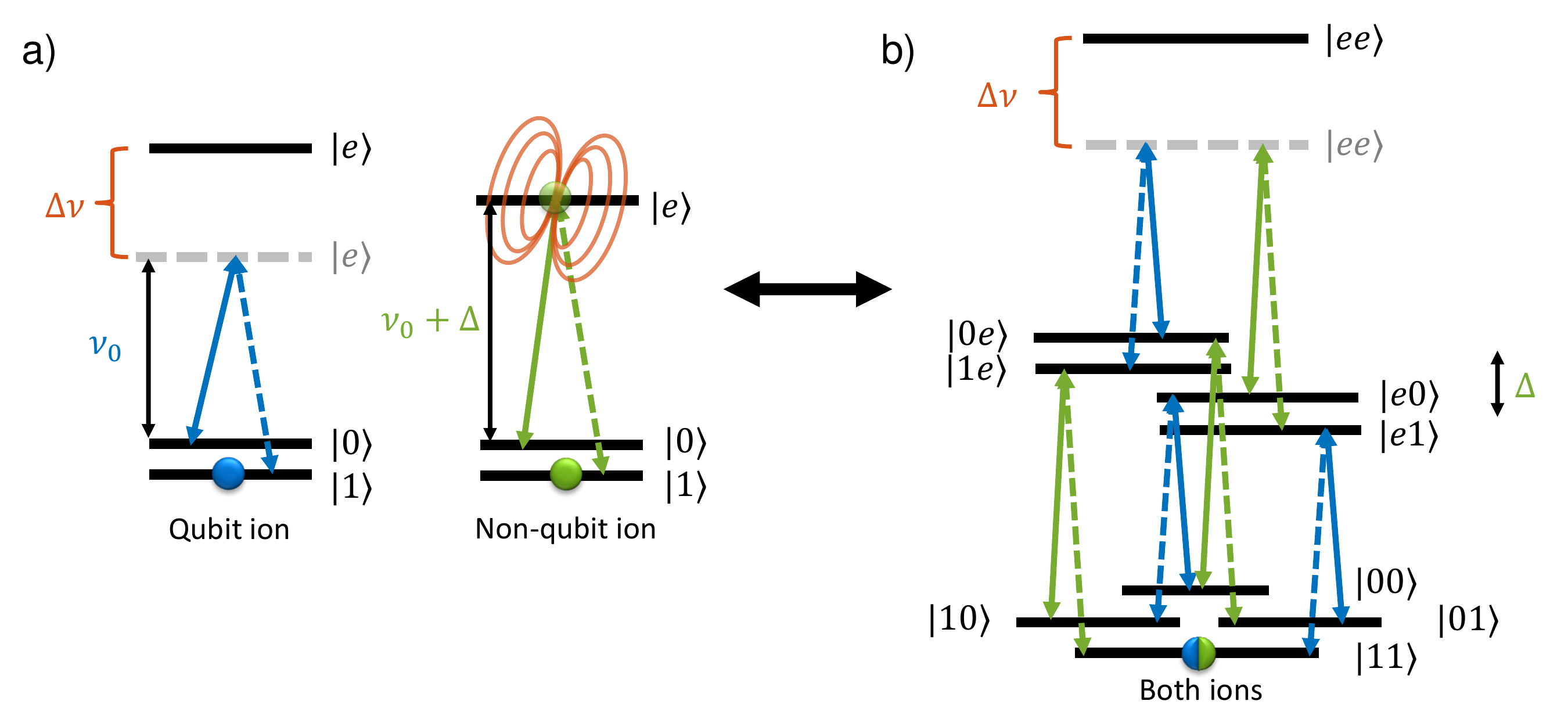}
\caption{\label{fig:Enery_levels}a) Shows the energy level structure for a qubit (blue) and a non-qubit (green) ion, including the two-color pulses driving each of the two ions. The ions can be different elements, but we assume that they are the same. However, the optical transitions of the non-qubit ion are detuned from the qubit by $\Delta$. When either of the two ions are excited, the other experiences a frequency shift, $\Delta\nu$, of its optical transitions. The fields driving the qubit can also be the fields driving the non-qubit, but they drive the transitions off-resonantly with a detuning of $\Delta$. Therefore, the coloring of the fields should mostly be used as a visual aid to easier identify which ion is driven on the various transitions. b) Shows the equivalent two-qubit energy level structure for the two ions. Note that the shift $\Delta\nu$ of the $|ee\rangle$ two-qubit state is a permanent energy level shift in this description.}
\end{figure*}

A Y$_2$SiO$_5$ crystal (or nanocrystal or thin film) is randomly doped with $^{153}$Eu dopants at the percent level. Two ground states of a single Eu ion can form a qubit, and gate operations can be performed via an excited state. At high concentrations qubits can be spaced at nanometer separation in three dimensions, thus providing very high qubit densities. The tight spacing also allows for strong dipole-dipole interactions between many nearby qubits, which is very beneficial as it provides a method to perform two-qubit or multi-qubit gate operations \cite{Ohlsson2002, Nilsson2002, Longdell2004a, Longdell2004, Ahlefeldt2013b}, and leads to high connectivity between qubits in rare-earth quantum computers \cite{Kinos2021c}. 

The linewidth of the optical transitions are roughly a kHz, and since the surrounding of each dopant is slightly different, there is an inhomogeneous broadening causing different ions to absorb at different frequencies. This broadening can be in the order of $100$ GHz \cite{Konz2003}. Thus, the laser used to control a qubit only interacts with a small subset of all dopants. Despite this, there could still be several thousands of non-qubit Eu ions within the focus size and bandwidth of the laser. When those non-qubit ions are excited, dipole-dipole interactions with the qubit ion can cause ISD, thus reducing the fidelity of gate operations performed on the qubit. To reduce this risk, optical pumping techniques \cite{Nilsson2002, Nilsson2004, Rippe2005, Lauritzen2012} and more complicated procedures \cite{Wesenberg2007} can be utilized. In Secs. \ref{sec:ISD_Crystal} and \ref{sec:ISD_crystal_analysis} we evaluate the ISD error such non-qubit dopants cause on a SQ gate operation. 

Lastly, the present work do not make any assumptions on how the state of a single-ion qubit is read out. However, a possible method is to use dedicated readout ions \cite{Wesenberg2007, Walther2015, Kinos2021}, which are co-doped with the qubit dopants but at a much lower concentration. 

\section{\label{sec:general_ISD}Microscopic treatment of instantaneous spectral diffusion}
ISD can occur when a non-qubit ion is excited and causes an unpredicted frequency shift, $\Delta\nu$, of the optical transitions of the qubit, as seen in Fig. \ref{fig:Enery_levels}a. This frequency shift leads to additional errors on gate operations performed on the qubit. The ISD can be simulated by constructing the Hamiltonian for the combined system as shown in Fig. \ref{fig:Enery_levels}b, where the excitation-dependent shift, $\Delta\nu$, comes in as a permanent shift of the $|ee\rangle$ energy level.

To separate the errors due to ISD from other error sources such as decay, decoherence, and internal crosstalk, we here investigate an idealized three-level lambda system where no decay or decoherence exist, and the pulses only drive the intended transitions, i.e., the pulse driving $|0\rangle\rightarrow|e\rangle$ only drives that transition, but does so for both the qubit and the non-qubit ion. SQ gate operations are performed using 2 two-color optical pulses resonant with the transitions $|0\rangle\rightarrow|e\rangle$ and $|1\rangle\rightarrow|e\rangle$, see more information in Appendices \ref{app:pulse_shape} and \ref{app:theory}, and Ref. \cite{Kinos2021a}. To focus our investigation on how ISD scales and can be understood, we only examine the case where a NOT operation is performed and all ions start in $|0\rangle+i|1\rangle$. This case was chosen since the NOT operation acts non-trivially on the initial state, but other cases would have been equally valid. If no ISD occurs in this idealized system, gate operations have no errors. 

\begin{figure}
\includegraphics[width=\columnwidth]{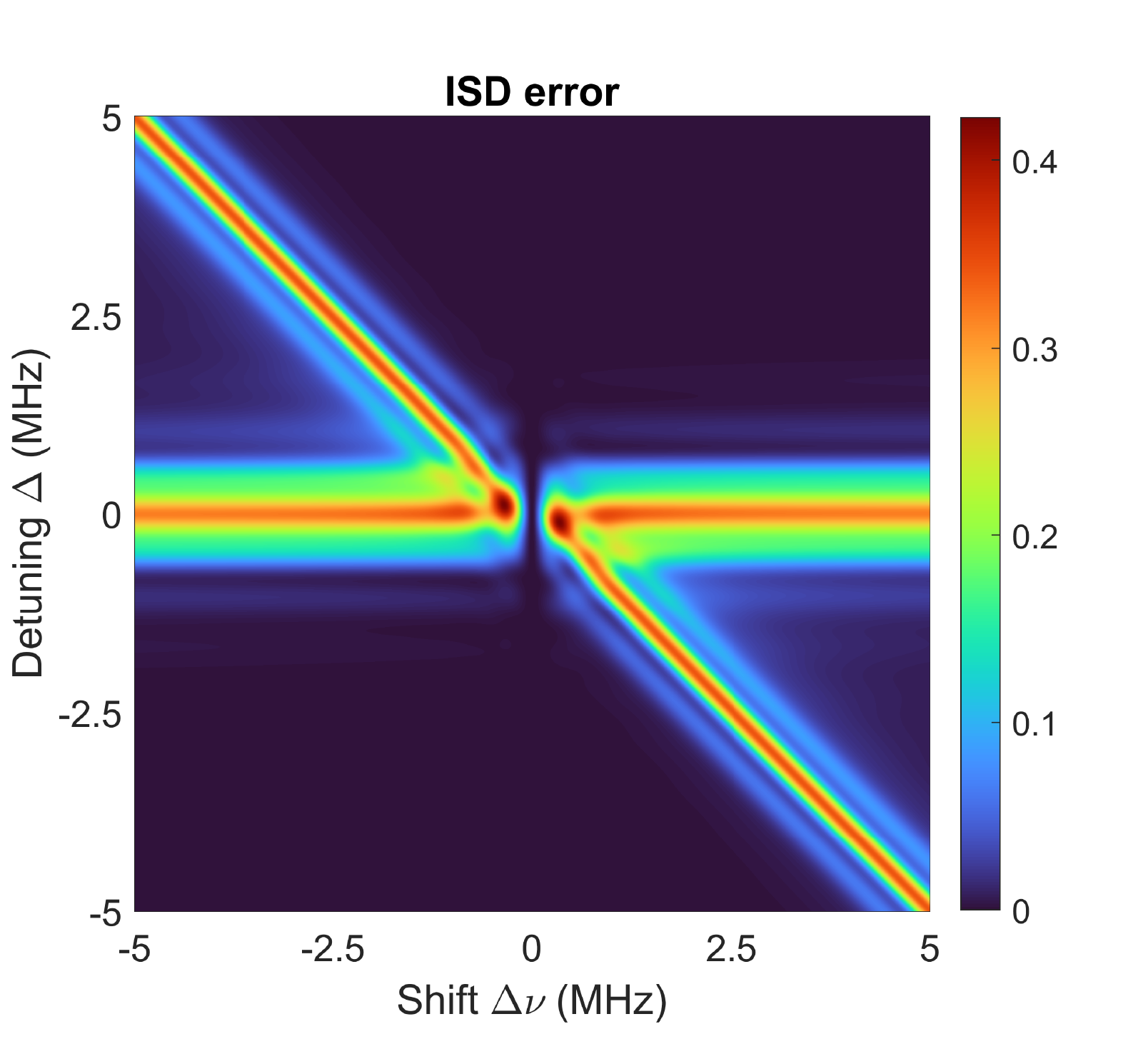}
\caption{\label{fig:Shift_detuning}The gate error due to ISD is seen as a function of both the shift $\Delta\nu$ and detuning $\Delta$ of one ion interacting with the qubit. Generally, there is an error if the non-qubit ion is excited and $\Delta\nu$ is non-negligible. When the detuning is larger than the gate bandwidth, the additional error is low for most shifts. However, when $\Delta\nu\approx-\Delta$ large errors still occur, since in these cases the non-qubit ion is shifted into resonance when the qubit is excited during the gate operation, thus affecting the evolution of the qubit. }
\end{figure}

The simulations in this article numerically solve the Lindblad master equation \cite{Manzano2020}, see Appendix \ref{app:simulations} for more information. The error of the qubit operation is calculated by first reducing the density matrix of the full system, $\rho_\text{full}$, which describes the qubit and all non-qubit ions, into the density matrix $\rho$ which only describes the qubit \cite{Nielsen2010};
\begin{equation}\label{eq:trace}
    \rho = \sum_{s}\langle I\otimes s|\rho_\text{full}|I\otimes s\rangle
\end{equation}

where $I$ is the identity matrix operating on our qubit, and the sum goes over all states $s$ of the non-qubit ions. The error of the operation, $\epsilon$, can then be calculated as
\begin{equation}\label{eq:error}
    \epsilon = 1 - \langle \Psi | \rho | \Psi \rangle
\end{equation}

where $\Psi$ is the target state of our qubit after the NOT operation has been applied, i.e., $\Psi = |0\rangle-i|1\rangle$. The additional SQ gate error due to ISD with one non-qubit ion is shown in Fig. \ref{fig:Shift_detuning} as a function of both the shift $\Delta\nu$ and the detuning $\Delta$ of the non-qubit ion. 

\begin{figure}
\includegraphics[width=\columnwidth]{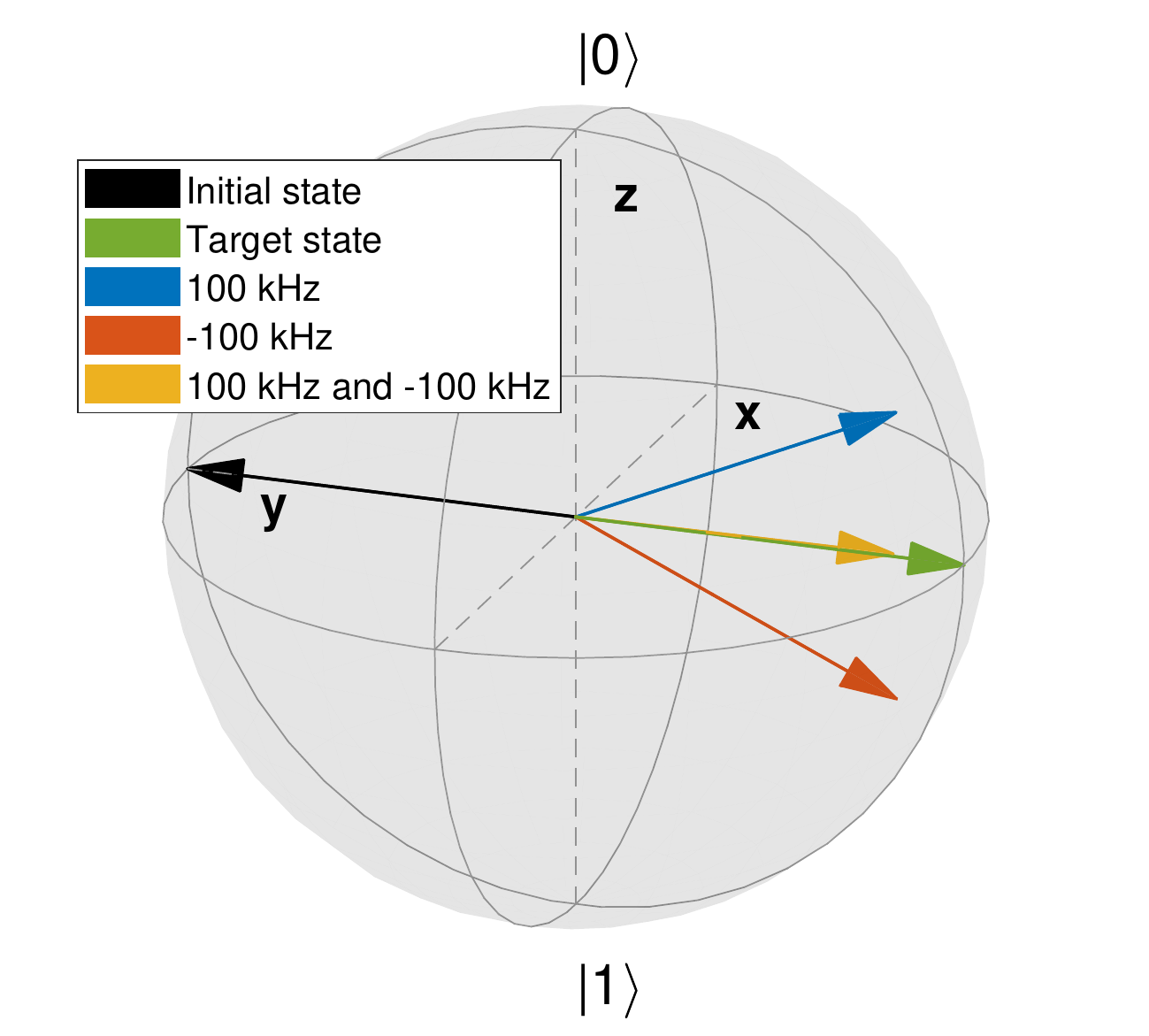}
\caption{\label{fig:BlochSphere}A NOT operation is performed on a qubit initially in $|0\rangle+i|1\rangle$ (black arrow) under the following circumstances: no ISD (green); ISD with one non-qubit ion with either $\Delta\nu_1 = 100$ kHz (blue) or $\Delta\nu_1 = -100$ kHz (red); and ISD with two non-qubit ions with $\Delta\nu_1 = 100$ kHz and $\Delta\nu_2 = -100$ kHz (yellow). The non-qubit ions are resonant with the qubit ($\Delta = 0$ MHz). The length of the qubit Bloch vector is reduced due to entanglement with the non-qubit ions. This error persists even though two non-qubit ions with opposite shifts interact simultaneously with the qubit (yellow). }
\end{figure}

The Bloch vector of the qubit, $\boldsymbol{a} = (u, v, w)$, is defined as;
\begin{equation}
    \begin{cases}
      u = \rho_{01} + \rho_{10} \\
      v = i(\rho_{01} - \rho_{10}) \\
      w = \rho_{00} - \rho_{11}
    \end{cases}
\end{equation}

where $\rho$ is the qubit density matrix obtained from Eq. \ref{eq:trace}. How this Bloch vector is altered when ISD occurs can be seen in Fig. \ref{fig:BlochSphere}. The qubit begins in $\boldsymbol{a} = (0, 1, 0)$ and the NOT operation ought to rotate this into $(0, -1, 0)$, which happens when no ISD is present. However, when $\Delta\nu\neq0$ the non-qubit ion affects the operation in two different ways. First, the Bloch vector changes direction and no longer solely has a $v$ component, i.e., the state vector is rotated. Second, the length of the Bloch vector, $|\boldsymbol{a}|$, is reduced as the qubit and the non-qubit ion become entangled. This is an unwanted entanglement since we do not keep track of the state and evolution of the non-qubit ion. The shrinkage of the Bloch vector occurs when we trace out the non-qubit system in order to examine only the qubit system as described in Eq. \ref{eq:trace}. Note that in the case when a qubit interacts with two non-qubit ions with opposite shifts, i.e., $\Delta\nu_2 = -\Delta\nu_1$, the two rotations mostly cancel each other, but the qubit still becomes entangled with the two non-qubit ions so an additional error still occurs. The effects of ISD are discussed further in Appendix \ref{app:ISD_theory}, where a theoretical approach is used.

\section{\label{sec:extrapolation_method}QBIES method - Qubit Bloch vector estimation based on independent error sources}
In general, a qubit Bloch vector can change in two ways: a rotation away from the target Bloch vector, and a reduction in the length of the Bloch vector. If $N$ different error sources exist, then the QBies method assumes that the rotations and shrinkages for different error sources are independent and estimates the qubit Bloch vector $\boldsymbol{a}$ in the following way: 
\begin{equation}\label{eq:extrapolation}
    \boldsymbol{a} = \prod_{n=1}^N (|\boldsymbol{a}_n|\boldsymbol{R}_n) \cdot \boldsymbol{a}_0
\end{equation}

where $\boldsymbol{a}_0$ is the Bloch vector obtained when no error is present, $|\boldsymbol{a}_n|$ and $\boldsymbol{R}_n$ are the length of the qubit Bloch vector and the rotation matrix required to turn the target state into the obtained Bloch vector, respectively, when the qubit is only disturbed by error source number $n$.

Generally, rotation matrices do not commute, so the order in which one applies them can matter. In this work, the rotations are applied starting with the ion causing the largest rotation before moving onto smaller rotations. Rotations do, however, commute if they rotate around the same axis, and in the case of the ISD investigated here the rotation axes are often quite similar. Furthermore, even for rotations around different axes the error due to them not commuting is small as long as the rotation angles are small. 

In the case of ISD, $N$ represents the number of non-qubit ions interacting with the qubit. Consequentially, error source number $n$ represents the ISD interaction between the qubit and only the $n^{\text{th}}$ non-qubit ion. How the results of the QBies method compares to running the full simulation in this case is discussed in Appendix \ref{app:validation}. Note that the full simulation of one qubit interacting with $N$ non-qubit ions requires a Hamiltonian containing $L^{N+1}$ energy levels, where $L$ is the number of energy levels per ion, whereas the QBies method can be performed using $N$ simulations of only $L^2$ energy levels since the effect of each non-qubit ion is treated separately. This means that the QBies method yields an exponential reduction in the number of energy levels required in the simulations. 

When studying ISD in Secs. \ref{sec:ISD_Crystal} and \ref{sec:ISD_crystal_analysis}, the QBies method relies on the assumption that the ISD errors from different non-qubit ions are independent and that we can neglect any interaction that might occur between different non-qubit ions. It also relies on the assumption that we can separate the effects of ISD from decay, decoherence, and internal crosstalk. Both of these assumptions are validated in Appendix \ref{app:validation}, where we conclude that the QBies method works really well in the vast majority of cases, but when the errors become large this method of estimating ISD becomes worse. If the errors are small, the non-qubit ions have low probabilities to be excited or they only interact weakly with the qubit. Therefore, in the vast majority of cases, the ISD error from one non-qubit ion is not significantly affected by the status of other non-qubit ions. Conversely, if the errors are large there is a higher probability that the non-qubit ions affect each other and the errors become dependent. In such cases, one might be forced to simulate the full system for the subset of interactions yielding a large error before applying the QBies method to the smaller errors. However, for applications concerning quantum computing the cases with small errors are the most relevant ones.

\section{\label{sec:ISD_Crystal}How to estimate the effect of instantaneous spectral diffusion}
\begin{figure*}
\includegraphics[width=\textwidth]{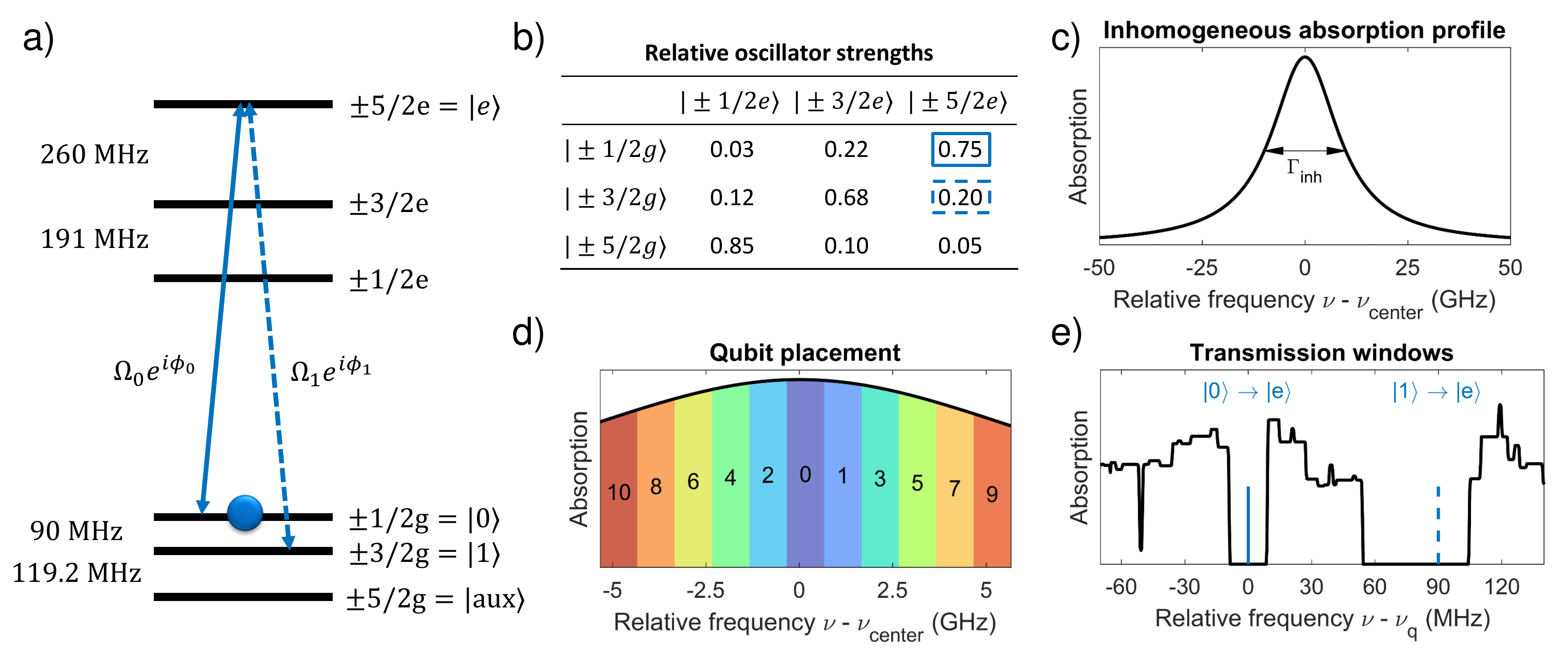}
\caption{\label{fig:Eu_Enery_levels}a) Energy level structure for $^{153}$Eu:Y$_2$SiO$_5$ site 1 \cite{Sun2005}, including the two optical driving fields with strengths $\Omega_0$ (solid line) and $\Omega_1$ (dashed line), and phases $\phi_0$ and $\phi_1$, respectively. b) The relative oscillator strengths of the different transitions \cite{Lauritzen2012}. c) The environment surrounding each dopant is slightly different which leads to inhomogeneities in the optical transitions of the dopants, as is indicated by the inhomogeneous absorption profile seen in the figure. The width of the profile, $\Gamma_\text{inh}$, is concentration dependent, see Eq. \ref{eq:Gamma_inh}, and can be up to hundred GHz broad \cite{Konz2003}. Here we use $c_\text{total} = 1\%$ as an example. d) We assume that each qubit reserves a frequency bandwidth of 1 GHz and label the qubits and their corresponding frequency channel as indicated in the figure. Qubit $q$ has a frequency channel that goes from $-335$ MHz to $665$ MHz relative to the $|0\rangle \rightarrow |e\rangle$ transition frequency, $\nu_q$, of the qubit. Frequency channel $q$ contains only qubit $q$, but it can contain many non-qubit ions, and these ions are assumed to only interact with the gate pulses performed on qubit $q$, since pulses controlling other qubits are assumed to be too far detuned. e) By using spectral hole burning and optical pumping techniques one can remove all non-qubit ions absorbing in the frequency regions around the two optical transitions used in a qubit. Here such zero absorption transmission windows with widths of roughly $18$ MHz and $50$ MHz, respectively, are shown as a function of the relative frequency $\nu - \nu_q$.}
\end{figure*}

This section explains how to estimate the effect of ISD on SQ gate operations due to dipole-dipole interactions with all randomly doped ions in the vicinity of the qubit. The interaction occurs since the static electric dipole moments of the ground and excited states are different, see Appendix \ref{app:dipole-dipole}, which is generally considered as the main contribution to ISD for non-Kramers dopants. In this work we examine the specific case of $^{153}$Eu:Y$_2$SiO$_5$ site 1 (europium doped into yttrium orthosilicate), whose properties can be found in Fig. \ref{fig:Eu_Enery_levels}a-b.

ISD is a stochastic phenomenon as the ISD error of a specific qubit strongly depends on the properties of the ions surrounding the qubit, e.g., their detunings and their dipole-dipole interactions with the qubit, as well as which ground states they initially reside in. Therefore, there is no single answer to how large the ISD error is for all qubits. Instead, we gather statistics to build a probability distribution over the risk that a single qubit suffers from ISD errors of different magnitudes. For each investigation we make, this is achieved by performing $1000$ simulations, where we in each case first randomize the properties of the ions surrounding a qubit and then estimate the ISD error they cause. Note that in a real crystal there can be correlations between the surroundings of different qubits if they are sufficiently close to each other. However, due to the strong spatial dependence of the dipole-dipole shift ($\Delta\nu \propto 1/|\boldsymbol{r}|^3$) and the fact that the different qubits have different transition frequencies, we argue that the probability that a certain ISD error occurs can be estimated using our method of studying independent qubit surroundings. The details of how the properties of the surrounding ions are randomized and how the ISD error is evaluated can be found in Appendices \ref{app:host_crystal} and \ref{app:interpolation_methods}, but the general idea is presented in the rest of this section. 

First, a fraction, $c_\text{total}$, of the yttrium ions within a sphere of $50$ nm radius centered around the qubit are replaced with Eu dopants, half of which are assumed to belong to site $1$. Note, that the crystals themselves do not need to be this small, instead this radius was picked because ions further away than this had a negligible contribution to the total ISD error, see Appendix \ref{app:ISD_N_and_r}. After doping the regions, the dipole-dipole shifts between any two ions can be calculated using Eq. \ref{eq:delta_nu} in Appendix \ref{app:dipole-dipole}. 

Second, each ion randomly obtains an optical transition frequency according to a Lorentzian line shape of the inhomogeneous absorption profile as seen in Fig. \ref{fig:Eu_Enery_levels}c. The full-width-at-half-maximum, $\Gamma_\text{inh}$, is assumed to grow linearly depending on the total doping concentration:
\begin{equation}\label{eq:Gamma_inh}
    \Gamma_\text{inh} = \Gamma_0 + \Gamma_c \cdot c_\text{total}
\end{equation}

where $\Gamma_0 = 1.8$ GHz is a concentration independent linewidth, $\Gamma_c = 1800$ GHz \cite{Sellars2004, Bengtsson2012}, and $c_\text{total}$ specifies the total atomic doping concentration between $0$ and $1$, where $1$ would be a fully doped stoichiometric crystal. Note, that this linear scaling is only valid for sufficiently low doping concentrations, but all concentrations used in this work ($c_{total} \leq 5\%$) fall into this regime. 

Following this, the $|0\rangle\rightarrow|e\rangle$ transition frequency of the qubit we investigate, from this point forward denoted by qubit index $0$, is set to be at the center of the inhomogeneous absorption profile. Furthermore, we assume that there exist other qubits and number them symmetrically growing outwards from this central frequency as is shown in Fig. \ref{fig:Eu_Enery_levels}d. 

ISD can be minimized by using spectral hole burning techniques \cite{Nilsson2002, Nilsson2004, Rippe2005, Lauritzen2012} to optically pump dopants between ground states so that the frequency regions close to the two qubit transitions $|0\rangle \rightarrow |e\rangle$ and $|1\rangle \rightarrow |e\rangle$ become free from absorption. Such semipermanent transmission windows can be seen in Fig. \ref{fig:Eu_Enery_levels}e, and the process to create them is described further in Appendix \ref{app:transmission_windows}. In order to create transmission windows for all qubits, each qubit must reserve a frequency bandwidth of roughly $1$ GHz, as indicated in Fig. \ref{fig:Eu_Enery_levels}d. 

In our model, all ions inside the reserved frequency range of a qubit only interact with the pulses controlling that qubit. Therefore, only the ions inside the reserved frequency range of qubit $0$ directly interacts with its pulses. However, other ions, which may be far detuned in frequency but spatially close to qubit $0$, can still cause ISD if they are partly excited before the gate operation on qubit $0$ is performed. Such ions can be excited when the qubits with indices $1$ to $Q$ perform $G$ gate operations, and in Sec. \ref{sec:ISD_crystal_analysis} the additional ISD error is studied as $Q$ and $G$ varies. 

Lastly, two different cases are studied in regards to the initial ground states of the ions. First, each ion is randomly placed in one of the three ground states with equal probabilities. Second, transmission windows are created for each qubit. Thus, the probability of which ground state an ion initially resides in depends on its detuning from its corresponding qubit. 

In summary, the ISD error of a single-ion qubit is estimated in the following way:
\begin{enumerate}[label*=\theenumi.]
  \item Create the qubit surrounding by randomly doping a limited spatial region around the qubit (we doped spherical regions with a radius of $50$ nm and concentrations ranging from $c_\text{total}=0.01\% \rightarrow 5\%$)
  \begin{enumerate}[label=\alph*.]
    \item Assign a position and static dipole moment direction to all ions, including the qubit
    \item Randomly assign optical transition frequencies to all ions, and set the $|0\rangle \rightarrow |e\rangle$ transition frequency of qubit $0$ to the center of the inhomogeneous absorption profile
    \item Determine the initial ground state for each non-qubit ion depending on whether transmission windows are used or not (in this work qubit $0$ starts in $|0\rangle+i|1\rangle$)
  \end{enumerate}
  \item Determine what rotation and shrinkage of the qubit $0$ Bloch vector each non-qubit ion causes due to ISD. Depending on if a non-qubit ion is inside or outside the frequency channel of qubit $0$, see Fig. \ref{fig:Eu_Enery_levels}d, different methods are used to determine these rotations and shrinkages, see more information in Appendix \ref{app:interpolation_methods}
  \item Use the QBies method described in Sec. \ref{sec:extrapolation_method} and Eq. \ref{eq:extrapolation} to estimate ISD due to all non-qubit ions or only a subset of them depending on what analysis is being made
\end{enumerate}

\section{\label{sec:ISD_crystal_analysis}The effect of instantaneous spectral diffusion on single-qubit gate operations}
This section studies how a gate operation is affected by ISD from non-qubit ions. The gate operations are performed using the pulses described in Appendix \ref{app:pulse_shape}, which are optimized to reduce the impact of ISD \cite{Kinos2021a}. If other gate parameters are used, the effect of ISD can be significantly worse compared to what is presented here. 

\begin{figure}
\includegraphics[width=\columnwidth]{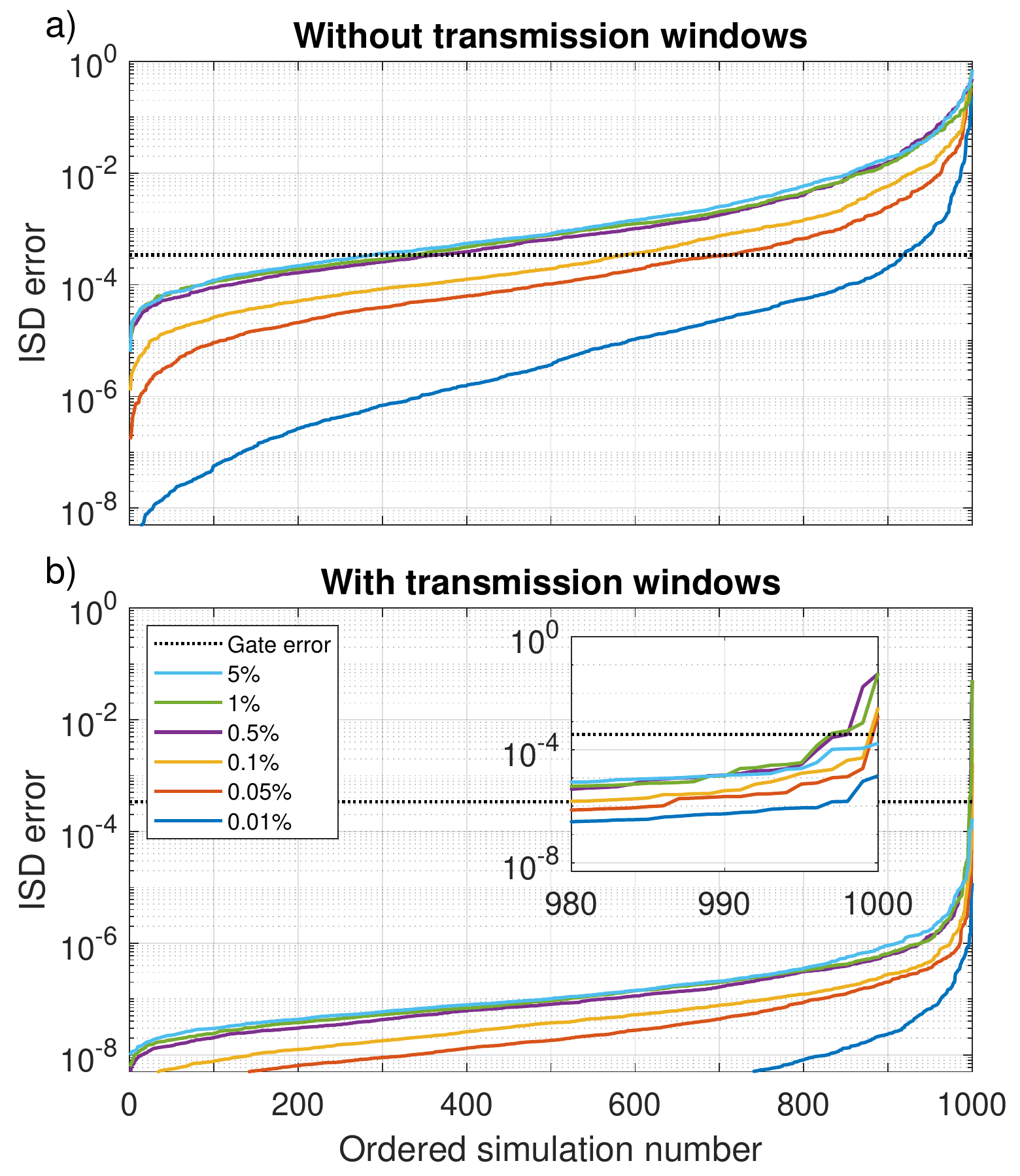}
\caption{\label{fig:ISD_concentration}The additional SQ gate error due to ISD caused by all non-qubit ions inside the reserved frequency range of qubit $0$ is estimated by performing $1000$ different simulations (horizontal axes) for each investigation we perform. Each simulation starts with a single qubit and estimates the ISD error by following steps 1-3 presented in the list in Sec. \ref{sec:ISD_Crystal}. The simulations are then ordered after the magnitude of the ISD error obtained. In a) no transmission windows are prepared, i.e., the non-qubit ions are equally likely to start in any of the three ground states. In b) the transmission windows shown in Fig. \ref{fig:Eu_Enery_levels}e are used, i.e., the probability that a non-qubit ion starts in a specific ground state depends on its detuning from the qubit. The inset zooms in on the simulations with the highest error. Several different doping concentrations are investigated, and both figures use the concentration labels shown in b). The dotted black lines show the SQ gate error due to decay, decoherence, and internal crosstalk for a NOT operation when no ISD is present.}
\end{figure}

The first case we investigate is the dependence on doping concentration when no transmission windows are created. Furthermore, no ions are excited before the qubit operation on qubit $0$ is attempted. Thus, only ions within the reserved frequency range of qubit $0$ cause ISD. The results are shown in Fig. \ref{fig:ISD_concentration}a. For all doping concentrations except the lowest, the majority of the simulations result in an additional ISD error that is at least in the same order of magnitude as the SQ gate error obtained when only considering decay, decoherence, and internal crosstalk. Initially, the error grows rapidly with increasing concentration. However, above a critical concentration, about $0.5\%$ for Eu:Y$_2$SiO$_5$, the increase in error slows down since instead of adding significantly more dopants per frequency channel as the concentration increases, the width of the inhomogeneous absorption profile mostly broadens, see Eq. \ref{eq:Gamma_inh}. Finally, note that these estimates of ISD are performed in the center of the inhomogeneous absorption profile and the effect is smaller in the wings. 

Fig. \ref{fig:ISD_concentration}b studies the same case except now transmission windows are used. Once more the concentration only matters below a critical value. However, ISD is now heavily suppressed thanks to the isolation of the qubit ion in frequency space. Therefore, we conclude that the usage of transmission windows is very important to limit the additional error due to ISD. 

\begin{figure*}
\includegraphics[width=\textwidth]{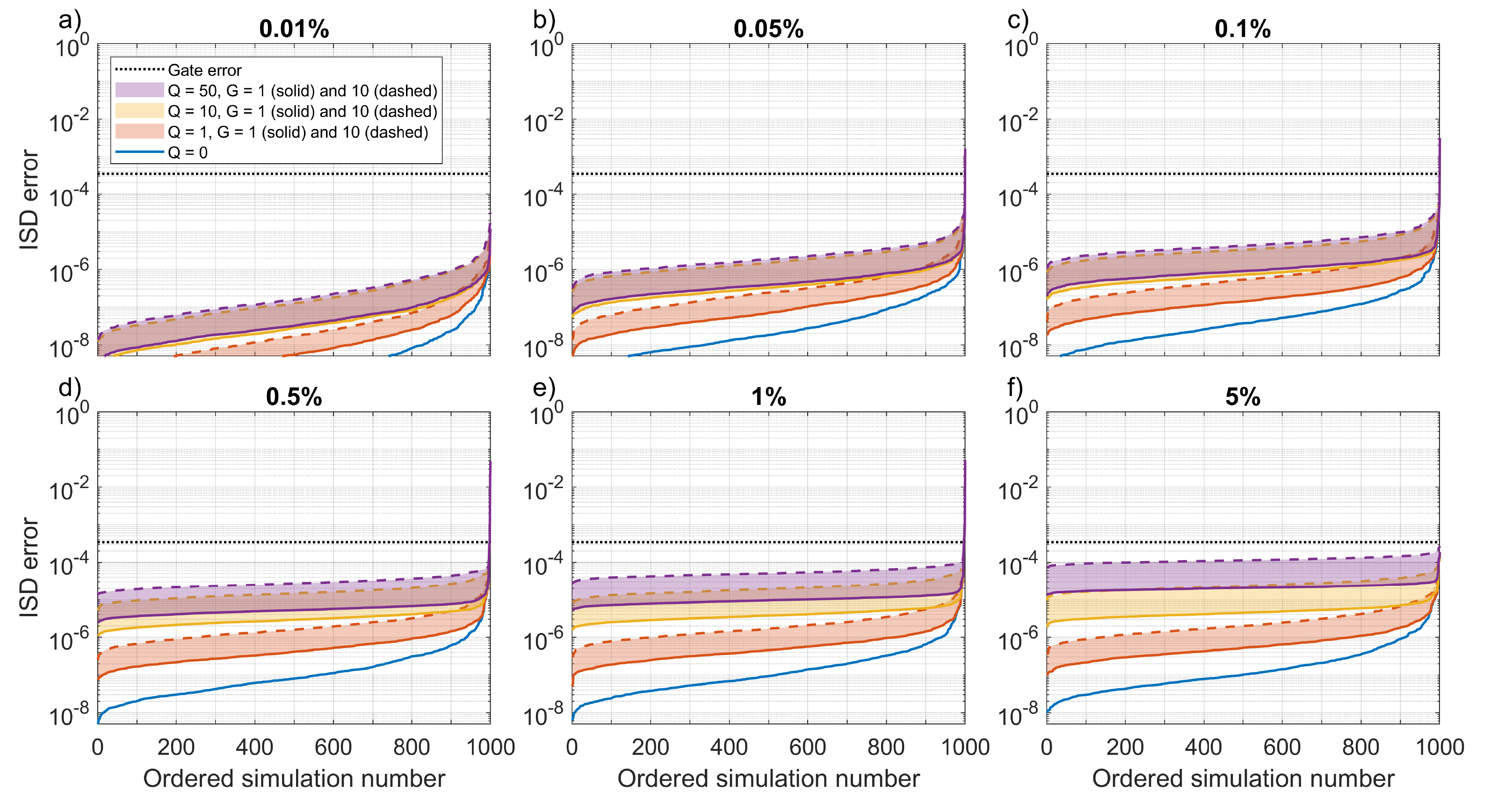}
\caption{\label{fig:ISD_q_and_op}Shows the ISD error as a function of the ordered simulation number (see the caption of Fig. \ref{fig:ISD_concentration} for more information). The total doping concentration $c_\text{total}$ varies from $0.01\% \rightarrow 5\%$ in figures a)-f). In all cases, two transmission windows (shown in Fig. \ref{fig:Eu_Enery_levels}e) are created for each qubit. The solid blue lines show the same results as Fig. \ref{fig:ISD_concentration}b, i.e., no gate operations were performed before the NOT operation on qubit $0$ was attempted. The other colors indicate how many other qubits, with indices $q = 1\rightarrow Q$, have undergone $G$ NOT operations before the NOT operation on qubit $0$ was attempted. The qubits with low indices are closest to the center of the inhomogeneous absorption profile as shown in Fig. \ref{fig:Eu_Enery_levels}d. In our simulations, $G$ is either $1$ (solid lines) or $10$ (dashed lines), and the colored regions span the interval between performing $1$ and $10$ NOT operations on each additional qubit.}
\end{figure*}

Furthermore, in Fig. \ref{fig:ISD_concentration}b only about $0.5\%$ (or $0.3\%$) of all simulations had qubits where the ISD errors were larger than $10\%$ (or $100\%$) of the normal SQ gate error. However, these estimates assume that all ions begin in one of their ground states, and thus only ions within the reserved frequency range of qubit $0$ cause ISD. Therefore, this corresponds to the expected effect of ISD when running the first gate operation on the first qubit in the quantum computer. When running subsequent gate operations, even if those are on other qubits and even if the operations run sequentially, the assumption that all ions begin in one of their ground states is no longer true. In Fig. \ref{fig:ISD_q_and_op} the ISD error on qubit $0$ is studied when up to $10$ NOT operations are performed on up to $50$ different qubits (i.e., $500$ operations in total) before the NOT operation on qubit $0$ is attempted. The general conclusions are presented here and a more detailed analysis is found in Appendix \ref{app:ISD_detailed_analysis}.

In the worst case studied here, $5\%$ doping concentration and running $10$ gate operations on $50$ additional qubits, the additional error due to ISD in the vast majority of the simulations is still below the SQ gate error obtained from the other error sources of decay, decoherence, and internal crosstalk. But it does increase the gate error by roughly $30\%$ to $60\%$. Thus, quantum computing using randomly doped rare-earth crystals is still feasible, despite all the numerous non-qubit ions in the vicinity of the qubits. However, it is again important to note that the gate operation pulses studied here are designed to also minimize the risk of ISD occurring \cite{Kinos2021a}. Hence, the ISD errors being lower than errors from all other sources is heavily dependent on the limited frequency bandwidth used by the gate operation pulses studied here and is not a general conclusion. 

For high doping concentrations, the ISD error increases by roughly $4\cdot10^{-7}$ for each gate operation that is performed on another qubit before the gate operation on qubit $0$ is attempted. Hence, the error scales linearly with the total number of gate operations. Furthermore, the error per gate can be reduced by broadening the inhomogeneous absorption profile without increasing the doping concentration, e.g., by co-doping with another rare-earth species \cite{Bottger2008}. Therefore, the ISD errors can be reduced even further compared to the results shown in Fig. \ref{fig:ISD_q_and_op}. 

We now turn to investigate how the initialization of the qubit ions might affect the ISD error. After transmission windows have been prepared, the qubit ion is in the $|\text{aux}\rangle$ ground state and must be transferred to either $|0\rangle$ or $|1\rangle$. This is done using resonant optical pulses, but in this process some non-qubit ions can also be transferred, thus creating a small ensemble peak of ions with absorption inside the transmission windows. These ions would interact strongly with the gate operation pulses, and thus most probably be excited during gate operations. However, they still need to spatially lie sufficiently close to the qubit in order to cause any significant ISD. The width of this peak, and thus how many such non-qubit ions are transferred, depend on the frequency width of the pulses used to perform the transfer. For more information about how these initialization pulses are performed, see Appendix \ref{app:transmission_windows}. Fig. \ref{fig:ISD_peak} shows how the ISD error changes as a function of the frequency width of these pulses. As can be seen, the non-qubit ions in the ensemble peak can cause a significant additional error when compared to the $0$ kHz case where it is assumed that only the qubit ion is transferred. Fortunately, there are schemes to remove such non-qubit ions \cite{Wesenberg2007}. Furthermore, if the ions are sufficiently spread out in frequency space, one alternative is to use transfer pulses with narrow frequency bandwidths such that only the qubit ion is transferred. 

\begin{figure}
\includegraphics[width=\columnwidth]{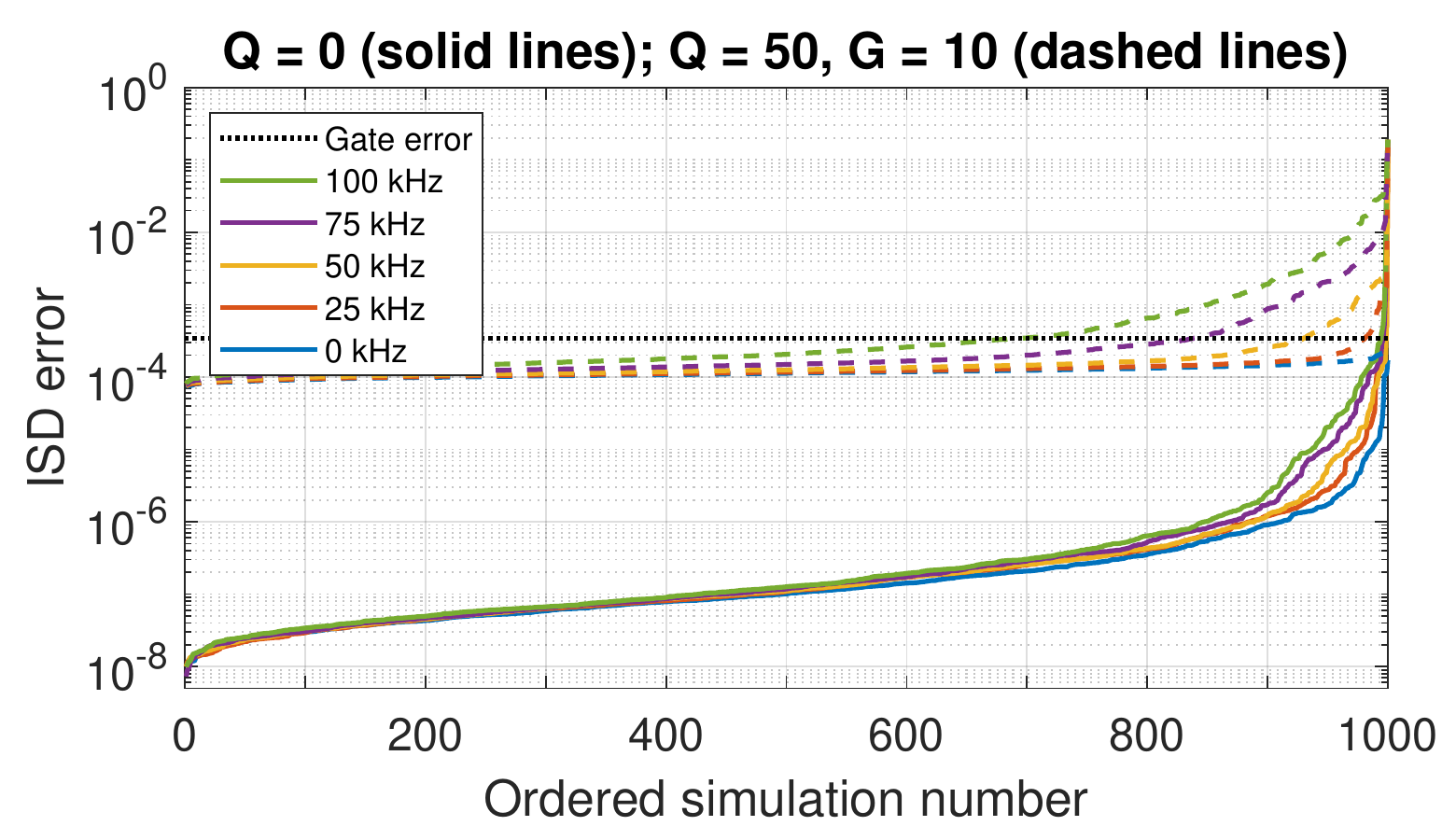}
\caption{\label{fig:ISD_peak}The ISD error as a function of the ordered simulation number when using a doping concentration of $5\%$ is shown for a few different frequency widths of the qubit initialization pulses (colors). In the $0$ kHz case only the qubit ion is transferred by the initialization pulses. In the solid lines no gate operations were performed on other qubits before the NOT operation on qubit $0$ was attempted. In contrast, the dashed lines show the results when $10$ NOT operations were performed on each of the qubits labeled $1\rightarrow 50$ before attempting the NOT operation on qubit $0$.}
\end{figure}

\section{\label{sec:conc}Conclusion}
In order to investigate how ISD affects gate operations on single-ion qubits in rare-earth crystals, where the stochastic behavior of the phenomenon is important, we have presented a microscopic treatment of ISD by modeling the dipole-dipole interactions between ions. 

In order to avoid the exponential scaling of the system size when examining ISD due to many non-qubit ions, we introduced the QBies method which can be used to estimate the total error from many error sources. The method is based on simulations only including one error source at a time, and works best when the errors are independent and small. 

ISD is then investigated under various conditions. It is concluded that transmission windows covering the two qubit transitions $|0\rangle\rightarrow|e\rangle$ and $|1\rangle\rightarrow|e\rangle$ are necessary in order to suppress the ISD errors. When using transmission windows, only about $0.3\%$ of the qubits have an additional error due to ISD that is larger than the error from other sources. However, for the majority of the qubits the ISD errors will be much lower. Thus, it is possible to construct gate operations with SQ gate errors of roughly $3 \cdot 10^{-4}$ to $5 \cdot 10^{-4}$ even when the additional error due to ISD is considered. Furthermore, the upper bound occurs only for the highest doping concentrations when several hundred gate operations have already been applied to other qubits in the quantum computer. Additionally, we discuss a way to further reduce the effect of ISD by reducing the number of non-qubit ions per frequency channel, e.g., by co-doping the crystal with another dopant to broaden the inhomogeneous absorption profile. 

The effect of the qubit initialization pulses is also studied. It is important to either perform such pulses in a precise way to reduce the number of non-qubit ions that are unintentionally transferred into the transmission windows when the qubit is initialized, or alternatively clean up the non-qubit ions after they have been transferred, to prevent additional ISD errors from such non-qubit ions. 

In summary, the current investigations show that it is possible to perform NISQ algorithms in randomly doped rare-earth crystals. Still, to determine the exact effect of ISD in the long-term usage of the quantum computer a more detailed analysis is required, but the current work provides a solid foundation to build upon. Furthermore, even if some qubits, due to the stochastic nature of the interaction, exhibit large ISD errors, one can choose not to use those qubit ions and therefore free up their frequency channels to be used by other qubit ions which have smaller ISD errors.

\begin{acknowledgments}
We want to thank Philippe Goldner for his input on the spectral and crystal site discussion in Appendix \ref{app:host_crystal}. This research was supported by Swedish Research Council (no. 2016-05121, no. 2015-03989, no. 2016-04375, and 2019-04949), the Knut and Alice Wallenberg Foundation (KAW 2016.0081), the Wallenberg Center for Quantum Technology (WACQT) funded by The Knut and Alice Wallenberg Foundation (KAW 2017.0449), and the European Union FETFLAG program, Grant No.820391 (SQUARE).
\end{acknowledgments}

\appendix
\section{\label{app:simulations}Information about the simulations}
All simulations where the effect of ISD is analyzed were performed by evolving the Lindblad master equation \cite{Manzano2020} using MATLAB's explicit Runge-Kutta ode45 function \cite{Dormand1980, Shampine1997}. In all simulations the qubit starts in $|0\rangle+i|1\rangle$ and a NOT operation is attempted. When simulating the $^{153}$Eu:Y$_2$SiO$_5$ system described in Fig. \ref{fig:Eu_Enery_levels} we also assume a zero magnetic field in the sense that the hyperfine levels are doubly degenerate, i.e., $|$+1/2g$\rangle$ overlaps with $|$-1/2g$\rangle$ and are therefore treated as one single level, $|\pm$1/2g$\rangle$. In most simulations studied here, decay, decoherence, and internal crosstalk effects are not included as those error sources are assumed to be independent from the ISD errors. One exception is when the SQ gate error due to only these sources and not ISD is evaluated. In that case the assumed optical life- and coherence times are $T_1 = 1.9$ ms and $T_2 = 2.6$ ms, respectively \cite{Equall1994}. Note that these values were obtained using a $10$ mT magnetic field but we do not include the small splitting this magnetic field inflicts on the otherwise degenerate energy levels. Finally, decay and decoherence processes between hyperfine states are always assumed to be negligible as they occur on time scales much longer than their optical counterparts \cite{Konz2003, Alexander2007, Arcangeli2014, Zhong2015}.

The Hamiltonians consisted of $L_\text{qubit}\cdot L_\text{ion}^{N}$ energy levels, where $L_\text{qubit}$ and $L_\text{ion}$ are the number of energy levels for the qubit and non-qubit ions, respectively, and $N$ is the number of non-qubit ions that cause ISD to the qubit. $L_\text{qubit}=3$ was always used for the qubit except for a single simulation when the SQ gate error due to decay, decoherence, and internal crosstalk was investigated where instead $L_\text{qubit}=6$ was used. The non-qubit ions used $L_\text{ion}=3$ in Sec. \ref{sec:general_ISD} where ISD was investigated in an idealized system. The simulations underlying the results in Sec. \ref{sec:ISD_crystal_analysis} used $L_\text{ion}=6$ for ions inside the reserved frequency range of qubit $0$ and $L_\text{ion}=2$ for ions outside. These simulations are described in Appendix \ref{app:interpolation_methods}.

\begin{figure}
\includegraphics[width=\columnwidth]{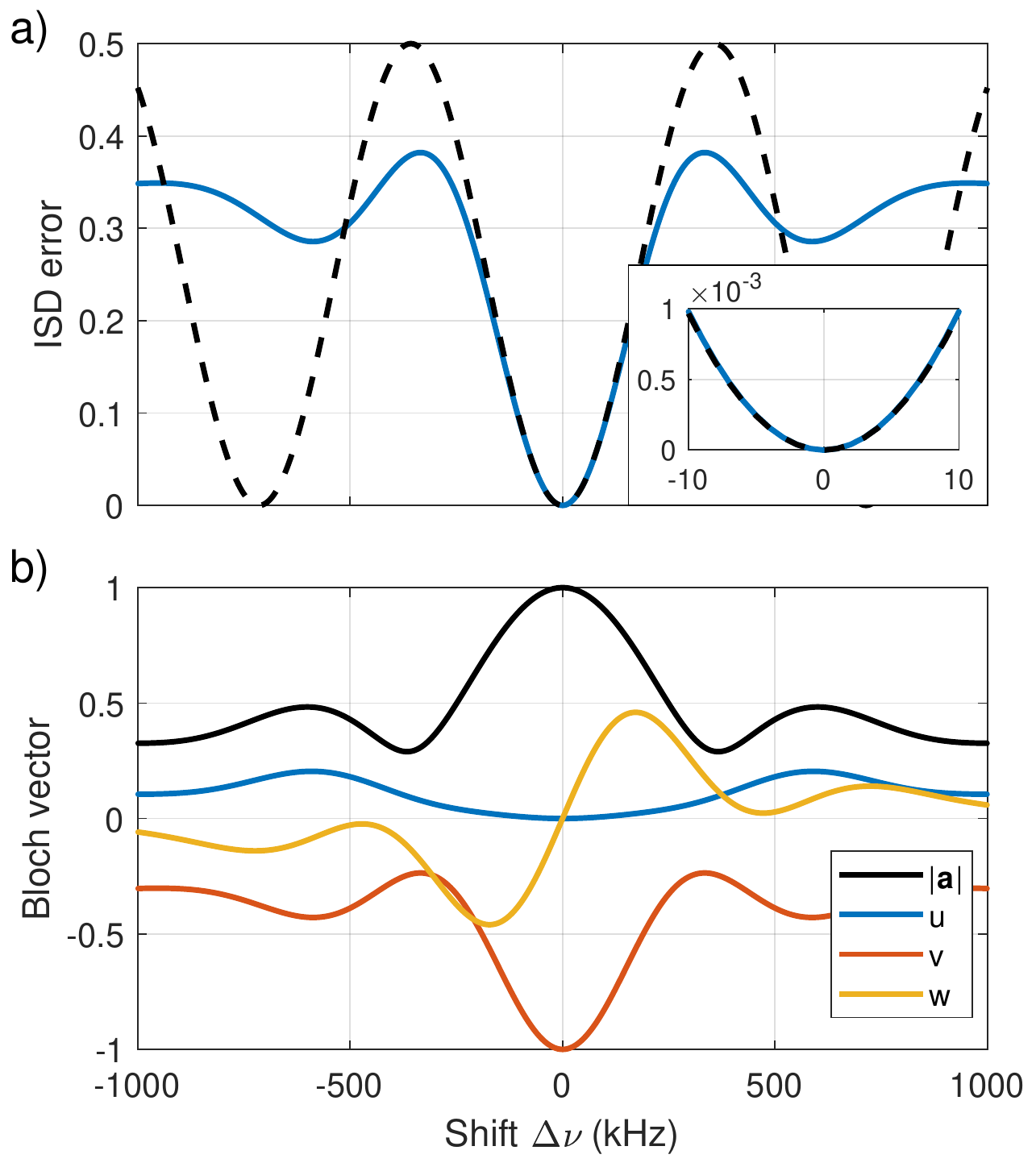}
\caption{\label{fig:Shift_1}a) Shows the ISD error as a function of the shift $\Delta\nu$ when the qubit interacts with one resonant ($\Delta = 0$) non-qubit ion (solid blue line). The theoretical error based on Eq. \ref{eq:theory_error}, where $t_e = 1.40$ $\upmu$s was optimized to give the best fit, is shown in the dashed black line. The theory assumes that both ions are excited despite the shift that occurs, i.e., it is only valid for small shifts, in this case $|\Delta\nu|$ less than roughly $100$ kHz. The inset zooms in around $\Delta\nu = 0$ kHz. b) Shows the length and components of the qubit Bloch vector $\boldsymbol{a} = (u, v, w)$ for the same simulation as in a).}
\end{figure}

\begin{figure*}
\includegraphics[width=\textwidth]{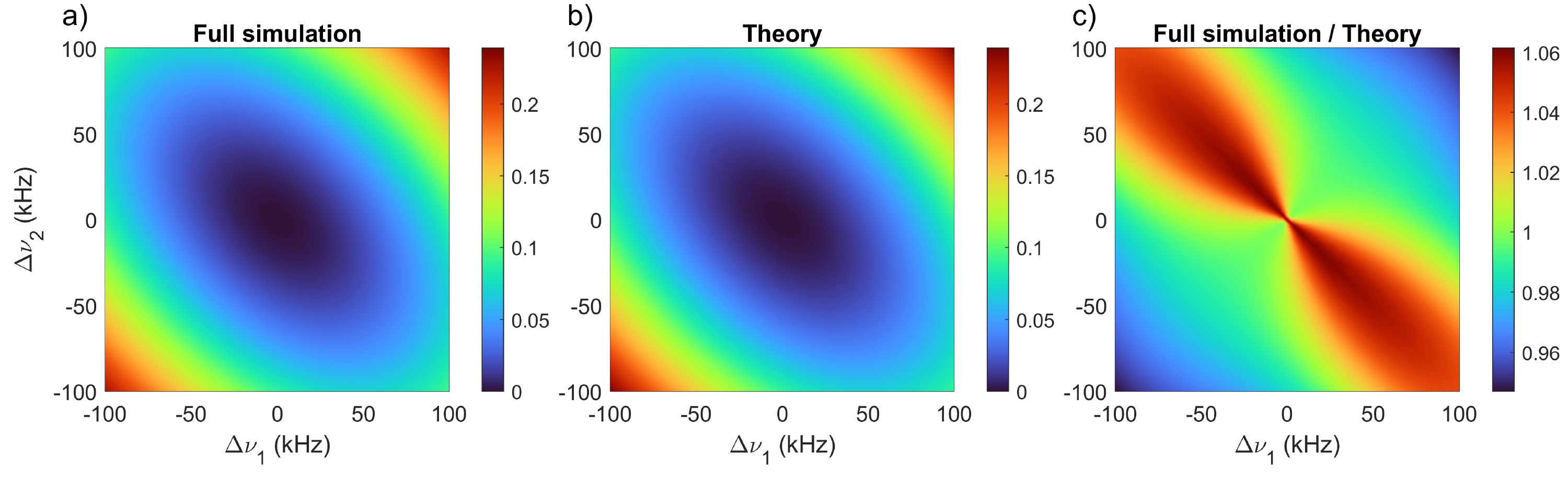}
\caption{\label{fig:Shift_2}Shows the ISD error from two non-qubit ions, both resonant with the qubit ($\Delta = 0$) but with varying shifts, $\Delta\nu_1$ and $\Delta\nu_2$, based on a) simulations, and b) the theoretical expression derived in Appendix \ref{app:theory}, where the duration $t_e = 1.40$ $\upmu$s was determined from the data of Fig. \ref{fig:Shift_1}. c) Shows the deviation between simulation and theory by dividing the ISD error obtained from the simulation by the error obtained from the theory.}
\end{figure*}

The relative and absolute tolerances for the ode45 MATLAB function were varied based on requirements: Figs. \ref{fig:Shift_detuning}, \ref{fig:BlochSphere}, and \ref{fig:Shift_1}-\ref{fig:Validation} used $10^{-10}$, whereas Figs. \ref{fig:Shift_detuning_map} and \ref{fig:ISD_excitation} used $3\cdot10^{-14}$. The global error of the simulation is estimated to be roughly $10$ times the local tolerances when running one SQ gate operation. The results of Figs. \ref{fig:ISD_concentration}-\ref{fig:ISD_peak} and \ref{fig:ISD_N_and_r}-\ref{fig:ISD_q_and_op_per_gate} are based on the simulations underlying the results shown in Figs. \ref{fig:Shift_detuning_map}, \ref{fig:ISD_excitation}, and \ref{fig:Holeburning}b. 

\section{\label{app:pulse_shape}Pulse shape of the SQ gate operations}
\begin{figure}
\includegraphics[width=\columnwidth]{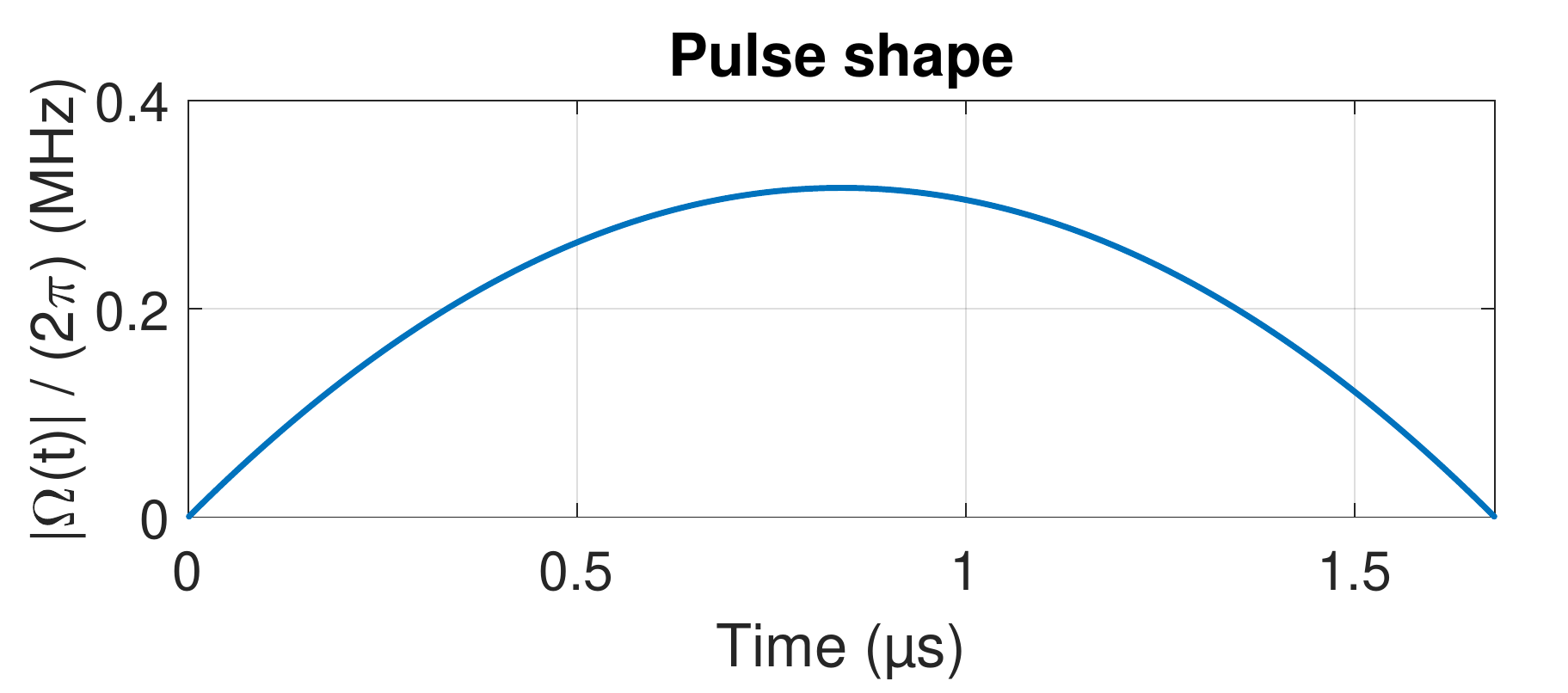}
\caption{\label{fig:pulse_shape}Shows the Rabi frequency envelope used for each of the two components of the two-color pulse. The pulse parameters in Eq. \ref{eq:Gaussian} are: $t_g = 1.68$ $\upmu$s and $\sigma = 4.16$ $\upmu$s.}
\end{figure}

The SQ gate operations used in this work are performed using 2 two-color optical pulses resonant with the transitions $|0\rangle\rightarrow|e\rangle$ and $|1\rangle\rightarrow|e\rangle$. The two driving fields have the same cut Gaussian pulse shape, $\Omega_0(t) = \Omega_1(t) = \Omega(t)$: 
\begin{equation}\label{eq:Gaussian}
  \Omega(t) =
    \begin{cases}
      C_1\cdot \text{exp}(-\frac{(t-t_g/2)^2}{2\sigma^2}) - C_2 & 0 \leq t \leq t_g\\
      0 & \text{otherwise}
    \end{cases}       
\end{equation}

where $t_g = 1.68$ $\upmu$s is the cut-off pulse duration, $\sigma = 4.16$ $\upmu$s is the standard deviation of the Gaussian, $C_1$ is chosen so that a pulse area of $\pi/\sqrt{2}$ is achieved, and $C_2$ enforces the shape to start and end at zero. These parameter values were optimized to achieve a low SQ gate error while also taking heed to minimize the risk of ISD occurring, as is discussed and motivated in Ref. \cite{Kinos2021a}. The pulse shape can be seen in Fig. \ref{fig:pulse_shape}. Since two pulses are required to perform a gate operation \cite{Kinos2021a} the total gate duration is $2t_g=3.36$ $\upmu$s.

Using a different pulse shape or different pulse parameters can affect the ISD errors. In Appendix \ref{app:ISD_theory}, we discuss how the ISD errors change as the pulse duration is altered. Furthermore, changing the pulse also affects errors from decay, decoherence, and internal crosstalk. Therefore, designing the optimal pulse is a complex optimization problem that is not analyzed further in this work. 

\section{\label{app:ISD_theory}ISD from an idealized theoretical point of view}
This section studies the effect of ISD under the assumption that all non-qubit ions are resonant with the qubit, i.e., $\Delta=0$ for all ions. This investigation is performed in order to build intuition in regards to ISD. First, the ISD error from one non-qubit ion as a function of the shift $\Delta\nu$ is investigated, see Fig. \ref{fig:Shift_1}a.

In general, the transitions to the $|ee\rangle$ two-qubit state are driven off-resonantly due to the shift, see Fig. \ref{fig:Enery_levels}b, resulting in phase and population errors on such transfers. However, when these errors are sufficiently low, i.e., the shift is much less than the frequency bandwidth of the pulse, one can assume that the pulses excite/deexcite both the qubit and the non-qubit ions regardless of the shifts that occur. Under these assumptions the interaction only comes in as a phase evolution on the shifted states, which in our case is the $|ee\rangle$ state. A theoretical expression for the ISD error when $N$ non-qubit ions interact with the qubit under these assumptions is derived in Appendix \ref{app:theory}. For the case of just one non-qubit ion, when both it and the qubit start in $|0\rangle+i|1\rangle$ and a NOT operation is performed, the error can be written as: 
\begin{equation}\label{eq:theory_error}
    \epsilon = \frac{1}{4} - \frac{1}{4}\cos(2\pi t_e \Delta\nu)
\end{equation}

where $t_e$ is the duration during which the two ions evolve their phase in the shifted $|ee\rangle$ two-qubit state, which is approximately equal to the pulse duration $t_g$, but also depends on the other gate parameters. The theoretical error is shown in the dashed black line of Fig. \ref{fig:Shift_1}a, where $t_e = 1.40$ $\upmu$s was optimized to give the best fit. 

As can be seen from the theoretical expression, a shorter pulse, yielding a shorter $t_e$, decreases the error due to ISD for a given shift since it spends less time in the $|ee\rangle$ state. However, one must also consider that if the shorter pulse requires a larger frequency bandwidth, then it might interact with more non-qubit ions, and although each individual ion might give a lower additional error, the combined effect of all non-qubit ions might not. Furthermore, an increased frequency bandwidth increases the risk of finding an ion that has a stronger interaction and thus shifts the qubit transitions more. 

A simple investigation can be made if we assume that the shifts occur due to dipole-dipole interactions, see Appendix \ref{app:dipole-dipole}, that the frequency bandwidth is inversely proportional to $t_e$, and that all frequency channels are equally likely to be populated by a non-qubit ion. If $t_e$ is decreased by a factor of $x$, then the bandwidth is increased by that same factor, as are the number of ions within the pulse bandwidth that may disturb the qubit. Because the ions are assumed to be randomly doped, the average distance from the closest non-qubit ion to the qubit scales as $|\boldsymbol{r}| \propto 1/x^{1/3}$. However, since the strength of the dipole-dipole shift scales as $1/|\boldsymbol{r}|^3$, the shift, $\Delta\nu$, increases by a factor of $x$. Therefore, provided that the assumptions are valid, to first order $t_e\Delta\nu$, and thus the additional error from the closest non-qubit ion, stays constant even if the pulse duration is changed. However, if the number of ions per frequency channel is not the same everywhere, then this is no longer true. For example, if transmission windows are used, then decreasing the pulse duration decreases the error due to ISD as long as the pulse is not interacting with the ions outside the transmission windows in a significant way. 

Let us now study the ISD error as a function of two non-qubit ions, both still resonant with the qubit and having shifts of $\Delta\nu_1$ and $\Delta\nu_2$, respectively. The results can be seen in Fig. \ref{fig:Shift_2}. If each ion interacts with the qubit alone they in general cause different rotations and length reductions of the qubit Bloch vector, see Fig. \ref{fig:Shift_1}b. Interestingly, when the shifts have opposite signs, the two rotations of the qubit Bloch vector originating from the interaction with each of the two ions partially cancel each other. If the shifts are completely opposite, i.e., $\Delta\nu_1 = -\Delta\nu_2$, the two rotations more or less cancel each other fully. However, in all cases the qubit still becomes entangled with the two non-qubit ions. Therefore, even in the case where the shifts are opposite an error still occurs. When the shifts have the same sign, the rotations from the two ions are in the same direction which gives an error that is larger than the sum of the individual errors originating from each ion interaction with the qubit alone. As can be seen in Fig. \ref{fig:Shift_2}b-c, the theory predicts the correct additional error due to ISD to within a few percent for these relatively small shifts.

\subsection{\label{app:theory}Theoretical error due to ISD for resonant ions with small shifts}
In this section we derive a theoretical expression of the additional SQ gate error due to ISD under the assumptions that the non-qubit ions are resonant with the qubit and that the gate operation pulses always excite both the qubit and all non-qubit ions regardless of the shifts that occur. This theoretical expression is therefore only valid for resonant cases and when the shifts are small compared to the frequency width over which the pulses can reliably transfer the ion from the ground to excited state and vice versa. The derivation in this section assumes that the gate operations are performed using $2$ two-color pulses \cite{Roos2004, Kinos2021a}, but it is possible to derive similar expressions for other gate protocols. 

When using two-color pulses, the qubit system has two superpositions, called bright and dark, which are coupled and uncoupled, respectively, to the excited state. These superpositions are defined as follows \cite{Roos2004}: 
\begin{eqnarray}\label{eq:BrightDark}
    |B\rangle = \frac{1}{\sqrt{2}} \left( |0\rangle + e^{-i\phi} |1\rangle \right) \nonumber\\
    |D\rangle = \frac{1}{\sqrt{2}} \left( |0\rangle - e^{-i\phi} |1\rangle \right)
\end{eqnarray}

where $\phi = \phi_1 - \phi_0$ is the relative phase between the two Rabi frequencies of the two-color pulse, see Fig. \ref{fig:Eu_Enery_levels}a. Both colors of the second two-color pulse have additional phases of $\phi_{0/1} = \phi_{0/1} + \pi - \theta$. The $2$ two-color pulses transfers the bright state to the excited state and back again with an additional phase of $e^{i\theta}$ while leaving the dark state unaffected. For a more detailed description see Ref. \cite{Kinos2021a}. 

The theoretical expression is derived in the bright/dark state basis and the core idea of the derivation is listed below: 
\begin{enumerate}
    \item The two-color pulses work as intended for both the qubit and the non-qubit ions, i.e., they transfer $|B\rangle\rightarrow|e\rangle$ and back again without errors
    \item Each state, e.g., $|BDD...DB\rangle$, acquires an additional phase of $e^{i\theta}$ for each $B$ component in the state after the operation is completed
    \item Each state, e.g., $|BDD...DB\rangle$, also acquires an additional phase due to ISD between the ions. This phase depend on how much the state $|eDD...De\rangle$ is shifted due to ISD, i.e., the initial state except all bright components have been excited
\end{enumerate}

To begin the derivation we list the initial state:
\begin{equation}
    |\Psi_{i}\rangle = \sum_{s=1}^{2^N} A_{Bs}|Bs\rangle + A_{Ds}|Ds\rangle
\end{equation}

where $A_{Bs}$ and $A_{Ds}$ are the coefficients for starting in state $|Bs\rangle$ and $|Ds\rangle$, respectively, with the $B$ and $D$ denoting the qubit state, and $s$ is the state of all $N$ non-qubit ions, which can be any of the $2^N$ combinations of them starting in $B$ or $D$. 

After the gate operation is performed the state is;
\begin{eqnarray}\label{eq:psi_f}
    |\Psi_{f}\rangle = \sum_{s=1}^{2^N} &&A_{Bs}e^{i\alpha(Bs)}|Bs\rangle + \nonumber\\
    &&A_{Ds}e^{i\alpha(Ds)}|Ds\rangle 
\end{eqnarray}

where $\alpha(x)$ is the phase;
\begin{equation}
    \alpha(x) = \theta\cdot n_B(x) - 2\pi t_e \sum_{i,j = \text{$B$ in $x$}} \Delta\nu_{ij}
\end{equation}

where $n_B(x)$ is the number of $B$ components in state $x$, $t_e$ is a duration which is proportional to the gate duration, $t_g$, and $\Delta\nu_{ij}$ is the shift between ions $i$ and $j$, measured in Hz. The sum goes over all combinations of ions $i,j$ that are $B$ in state $x$, e.g., state $x=|BDBBD\rangle$ results in the combinations $(i,j) = (0,2), (0,3)$, and $(2,3)$. 

In order to trace out the non-qubit ions we must first define the density matrix of the full system, $\rho_\text{full}$, and our qubit, $\rho$; 
\begin{eqnarray}
    \rho_\text{full} = |\Psi_f\rangle\langle\Psi_f| \\
    \rho = \sum_{s=1}^{2^N}\langle I\otimes s|\rho_\text{full}|I\otimes s\rangle
\end{eqnarray}

Using the expression for $|\Psi_f\rangle$ from Eq. \ref{eq:psi_f} we get;
\begin{eqnarray}
    \rho = \sum_{s=1}^{2^N} \left(A_{Bs}e^{i\alpha(Bs)}|B\rangle
    + A_{Ds}e^{i\alpha(Ds)}|D\rangle\right) \nonumber\\
    \cdot \left(A_{Bs}^*e^{-i\alpha(Bs)}\langle B| 
    + A_{Ds}^*e^{-i\alpha(Ds)}\langle D|\right) 
\end{eqnarray}

We can now transform back to the qubit system of $|0\rangle$ and $|1\rangle$ using Eq. \ref{eq:BrightDark} and then calculate the Bloch vector components; 
\begin{eqnarray}
    u =&& \rho_{01} + \rho_{10} \nonumber\\
    v =&& i(\rho_{01} - \rho_{10}) \nonumber\\
    w =&& \rho_{00} - \rho_{11}
\end{eqnarray}

This results in; 
\begin{eqnarray}\label{eq:theory_uvw}
    u =&& \sum_{s=1}^{2^N} \cos(\phi)(|A_{Bs}|^2 - |A_{Ds}|^2) \nonumber 
    + 2\sin(\phi)\text{Im}(\xi(s) ) \\
    v =&& \sum_{s=1}^{2^N} -\sin(\phi)(|A_{Bs}|^2 - |A_{Ds}|^2) \nonumber 
    + 2\cos(\phi)\text{Im}(\xi(s)) \\
    w =&& \sum_{s=1}^{2^N} 2\text{Re}(\xi(s))
\end{eqnarray}

where 
\begin{equation}
    \xi(s) = A_{Bs}A_{Ds}^*e^{i\beta(s)} 
\end{equation}

where $\beta(s) = \alpha(Bs) - \alpha(Ds)$, which can be simplified to; 
\begin{equation}
    \beta(s) = \theta - 2\pi t_e \sum_{j = \text{$B$ in $s$}} \Delta\nu_{0j}
\end{equation}

where the summation now only looks at the interactions between the qubit, with index $0$, and the non-qubit ions in state $|B\rangle$,  e.g., state $s=|DBBD\rangle$ results in the combinations $(0,j) = (0,2),$ and $(0,3)$, i.e., $\beta(s)$ does not include the interaction between non-qubit ions. Remember that state $s$ only includes the states of the non-qubit ions, whose first index is $1$. 

The expressions in Eq. \ref{eq:theory_uvw} work for any number of non-qubit ions, $N$, any initial state, $A_{Bs}$ and $A_{Ds}$, and arbitrary gate operations, $\phi$ and $\theta$. For the case of one non-qubit ion interacting with the qubit with a shift of $\Delta\nu$, when both ions start in $|0\rangle+i|1\rangle$ and a NOT operation, $\phi=\pi$ and $\theta=\pi$, is performed, the results simplifies to;
\begin{eqnarray}
    u =&& \ 0 \nonumber \\ 
    v =&& \ -\frac{1}{2} - \frac{1}{2}\cos(2\pi t_e \Delta\nu) \nonumber \\
    w =&& \ \frac{1}{2}\sin(2\pi t_e \Delta\nu)
\end{eqnarray}

Since the initial state has $v=1$ and a NOT operation is performed, the target state is $v=-1$ and the error of the operation can be calculated as $\epsilon = (1+v)/2$, i.e., 
\begin{equation}
    \epsilon = \frac{1}{4} - \frac{1}{4}\cos(2\pi t_e \Delta\nu)
\end{equation}

\section{\label{app:validation}Validating the QBies method}

\begin{figure*}
\includegraphics[width=\textwidth]{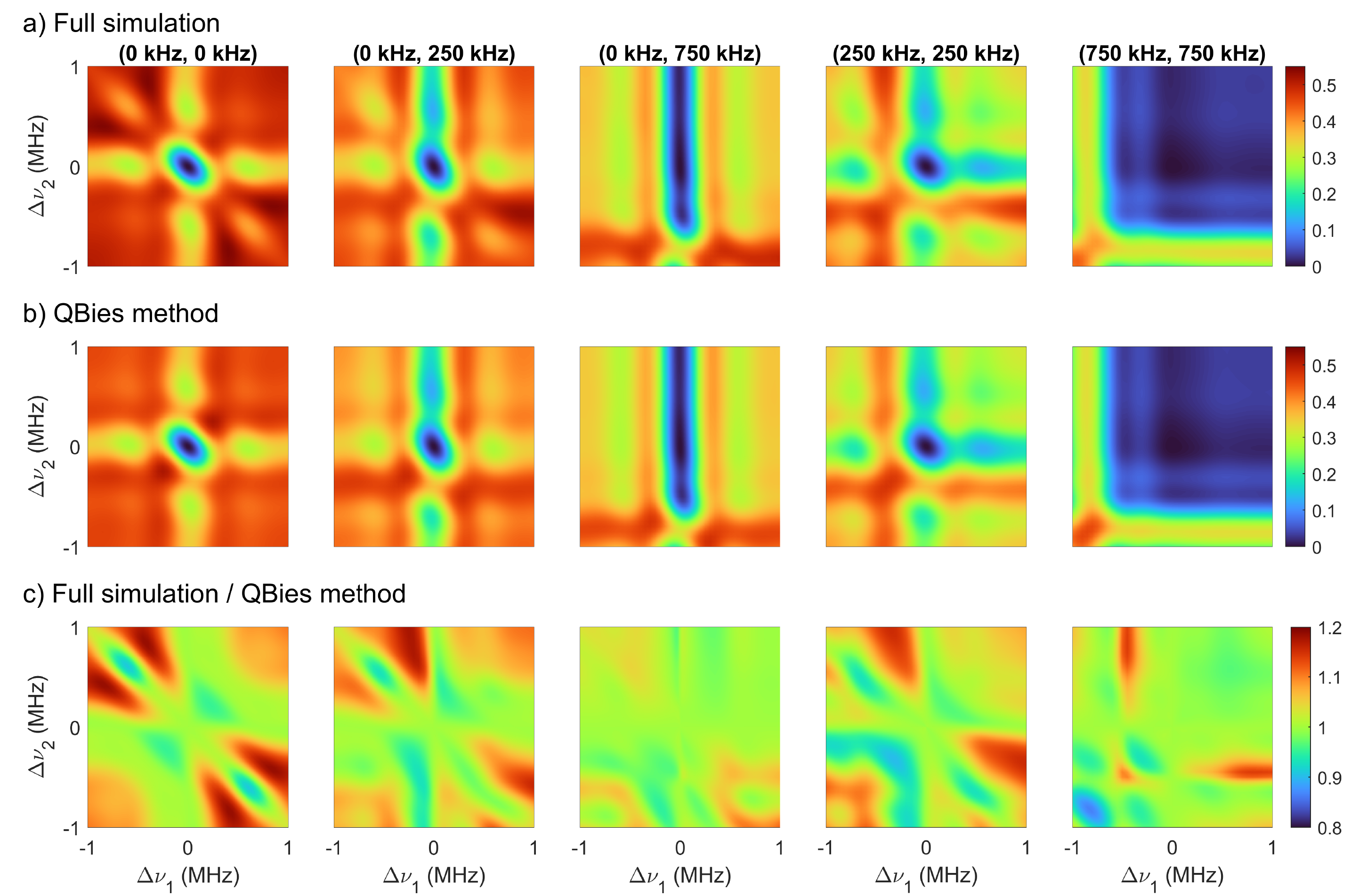}
\caption{\label{fig:Shift_detuning_extrapolation}The ISD error on a qubit due to interaction with two non-qubit ions is shown. The shifts, $\Delta\nu_1$ and $\Delta\nu_2$, are varied along the horizontal and vertical axes of each graph, respectively. The different columns show the results for different detunings $(\Delta_1$, $\Delta_2)$ of the two non-qubit ions, where a zero detuning means that the non-qubit ion is resonant with the qubit. The results are shown based on a) simulations of the full system including the qubit and the two non-qubit ions, and b) the QBies method presented in Eq. \ref{eq:extrapolation}, which is based on two separate simulations using only the qubit and a single non-qubit ion in each simulation. c) Shows the deviation between the simulation and the QBies method by dividing the simulation results by the QBies results.}
\end{figure*}

This section explores the validity of the QBies method presented in Eq. \ref{eq:extrapolation}. To start, Fig. \ref{fig:Shift_detuning_extrapolation} shows a comparison between running the full simulation and using the QBies method for a qubit interacting with two non-qubit ions. In all cases shown, the true additional error due to ISD is at most $\pm20\%$ compared to the error obtained using the QBies method. Furthermore, this ratio approaches $1$ when the errors are low, i.e., when the detunings of the non-qubit ions are large or when the shifts are small. This is good for the case of rare-earth quantum computing, since in a real crystal there are many more ions that are far detuned from the qubit or have weak interactions due to the $1/|\boldsymbol{r}|^3$ scaling of the dipole-dipole shift, see Appendix \ref{app:dipole-dipole}, and the fact that the number of ions at a certain distance $|\boldsymbol{r}|$ scales as $|\boldsymbol{r}|^2$. 

We now continue by investigating the following assumptions: 1) ISD errors can be separated from the other error sources of internal crosstalk and decay/decoherence, and 2) ISD errors from different non-qubit ions interacting with the qubit can be separated from each other, and shifts between different non-qubit ions can be neglected. 

These issues are investigated by first running full simulations where everything is included at once, and then comparing those results to when one uses the QBies method based on running simplified systems where the different error sources are separated. In these simulations a NOT operation is performed on the qubit ion, which begins in $|0\rangle+i|1\rangle$, and is described by a partly idealized system with three energy levels, $|0\rangle$, $|1\rangle$, and $|e\rangle$. We let the qubit interact with $N$ additional non-qubit ions, that each have a $50\%$ chance of being described by the same three-level system as the qubit and also interact with the gate pulses, and a $50\%$ chance of being described by a simplified two level system with only one ground and one excited state, not interact with the gate pulses, and instead have some initial population in the excited state. The detunings of the non-qubit ions (only relevant if they are interacting with the gate pulses), are randomized by a logarithmic uniform distribution between $1$ kHz and $100$ MHz, i.e., the randomized values are $10^x$ Hz, where $x$ is uniformly distributed between $3$ and $8$. Similarly, the shifts between the qubit and non-qubit ions, as well as between different non-qubit ions, are randomized by a logarithmic uniform distribution between $100$ Hz and $10$ MHz. The non-qubit ions that do not interact with the gate pulses have an initial excited state population that is randomized by another logarithmic uniform distribution now between $10^{-9} \rightarrow 10^{-4}$, where, e.g., $10^{-4}$ means that the ion starts in a mixed state with a $10^{-4}$ probability to be in the excited state. All these ranges were picked to validate the assumptions over a large range of different values, and to make sure that each individual variable could affect the results in a significant way, while simultaneously resulting in total errors that span a large range. In each investigation we perform $1000$ different simulations where all parameters are randomized again. 

\begin{figure*}
\includegraphics[width=\textwidth]{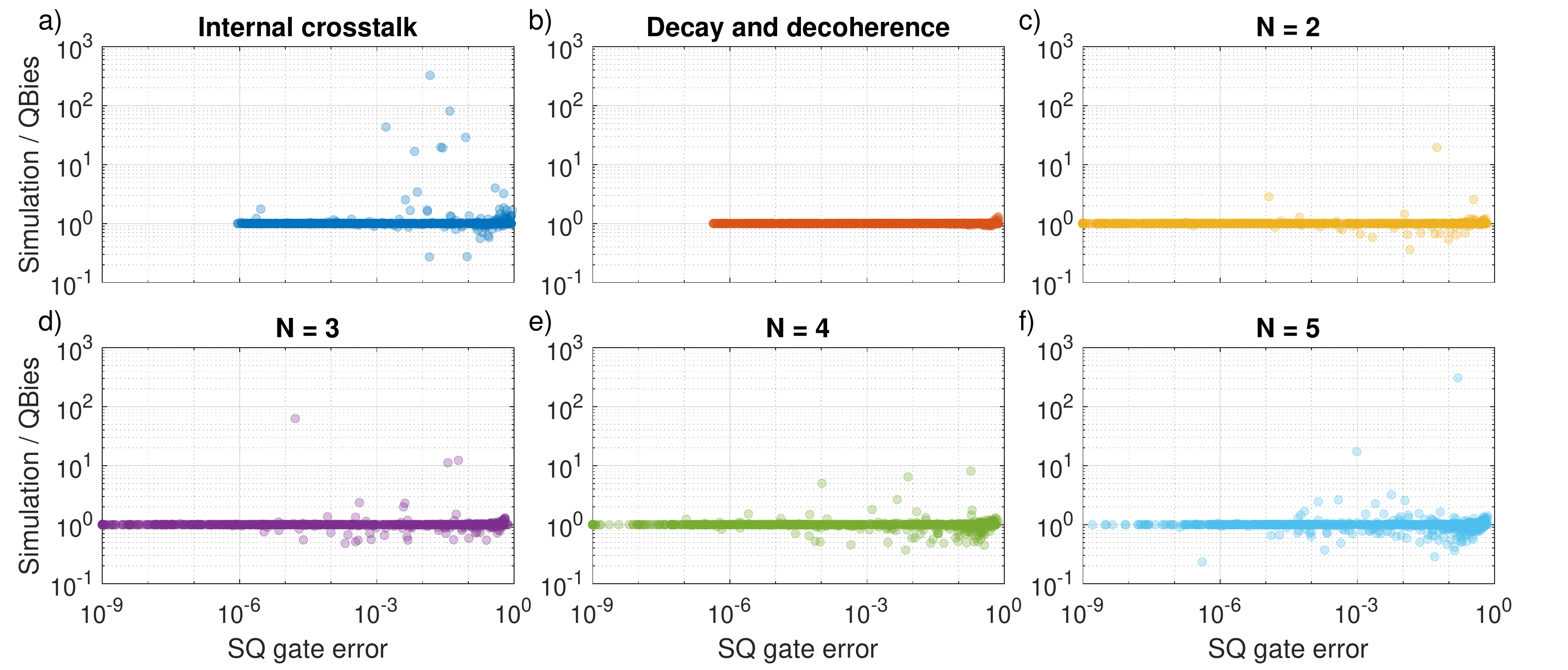}
\caption{\label{fig:Validation}This figure validates the assumption that in the vast majority of cases the rotation and shrinkage of the qubit Bloch vector caused by ISD from one non-qubit ion is independent from the rotations and shrinkages caused by internal crosstalk, decay and decoherence, and ISD from other non-qubit ions. The graphs show the ratio between the total error obtained when either running the full simulation of the entire system or using the QBies method described in Eq. \ref{eq:extrapolation} in order to estimate the total error. The horizontal axes show the total SQ gate error obtained in the full simulation. Six different investigations are performed: a) separating the errors of ISD from errors of internal crosstalk; b) separating the errors of ISD from errors of decay and decoherence; c-f) separating the errors of ISD from $N = 2\rightarrow5$ different non-qubit ions interacting with the qubit. In each investigation we simulate $1000$ different cases, shown as half-transparent circles. In all investigations the QBies method yields very similar results in the vast majority of cases when compared to running the full simulation, i.e., most circles lies close to $1$ on the vertical axes. Furthermore, most deviations occur when the total SQ gate error is high.}
\end{figure*}

First, the separation of ISD errors from errors due to internal crosstalk is investigated. Here the qubit only interacts with one non-qubit ion, and the partly idealized systems have no decay or decoherence. However, when running the simulation of the full system the gate operation pulses are allowed to drive all transitions, and the two ground states are separated by a frequency drawn from the logarithmic uniform distribution of $100$ kHz to $100$ MHz. Then a second simulation is performed with exactly the same randomized values, but where the gate operation pulses only drive the intended transitions. This simulation gives the rotation and shrinkage of the qubit Bloch vector due to errors of ISD only. A third simulation is then performed were the gate operation pulses once more interact with all transitions, but it only contains the qubit ion, i.e., there is no ISD. This last simulation provides a Bloch vector that only contains the errors due to internal crosstalk. Eq. \ref{eq:extrapolation} is now used to estimate the total error due to both ISD and internal crosstalk where the rotation and shrinkage are obtained from the second simulation listed above, and $\boldsymbol{a}_0$ is the Bloch vector obtained from the third simulation listed above. Finally, the total error obtained from the first full simulation due to both internal crosstalk and ISD is compared with the error estimated based on the last two simulations. This is repeated $1000$ times and the results can be seen in Fig. \ref{fig:Validation}a. As can be seen, the ratio of the error obtained from the full simulation divided by the error obtained from the QBies method is roughly equal to $1$ for the vast majority of cases. Furthermore, the deviations from this mostly occur when the total SQ gate error is high. 

A similar investigation is performed to validate the assumption that ISD errors can be separated from errors due to decay and decoherence. The procedure is similar to that listed above, except instead of driving multiple transitions decay and decoherence is included or not. To validate the assumption over a large range of errors the excited state lifetime, $T_1$, is randomized from a logarithmic uniform distribution between $100$ ns and $1$ s. The results can be seen in Fig. \ref{fig:Validation}b. Here the QBies method works even better and the difference between it and running the full simulation is negligible. 

Lastly, the assumption that ISD errors from $N = 2\rightarrow 5$ non-qubit ions can be separated from each other is investigated. There is no decay, decoherence, or internal crosstalk in any of these investigations, the results can be seen in Fig. \ref{fig:Validation}c-f, and the conclusions are similar as those written above. 

In summary, it is possible to separate different error sources such as internal crosstalk, decay and decoherence, and ISD originating from different non-qubit ions in the vast majority of cases, especially when the total SQ gate error is low. 

\bigskip
\bigskip

\section{\label{app:dipole-dipole}Dipole-dipole interactions}
The ISD considered in this article is modeled as a dipole-dipole interaction occurring since the static electric dipole moments of the ground and excited states, $\boldsymbol{\mu}_g$ and $\boldsymbol{\mu}_e$, are different. Thus, when an ion is either excited or deexcited, its charge distribution is modified and the resulting electric field change affects nearby ions. The frequency shift, $\Delta\nu$, on the optical transitions of such nearby ions due to this interaction can be calculated as follows \cite{Jackson1998};
\begin{eqnarray}\label{eq:delta_nu}
    \Delta\nu =&& \frac{k}{|\boldsymbol{r}|^3} (\Delta\boldsymbol{\mu}_A\cdot\Delta\boldsymbol{\mu}_B - 3(\Delta\boldsymbol{\mu}_A\cdot\hat{\boldsymbol{r}})(\hat{\boldsymbol{r}}\cdot\Delta\boldsymbol{\mu}_B)) \nonumber\\
    k =&& \frac{(\epsilon(0) + 2)^2}{9\epsilon(0)} \frac{1}{4\pi\varepsilon_0 h} 
\end{eqnarray}

where $\boldsymbol{r}$ is the spatial vector pointing from ion $B$ to ion $A$, $\hat{\boldsymbol{r}}$ is the normalized spatial vector, and $\Delta\boldsymbol{\mu_}{A/B}$ is the difference $\boldsymbol{\mu}_g - \boldsymbol{\mu_}e$ for ions $A$ and $B$, respectively. The first term in the constant $k$ is a local field correction due to the crystal \cite{Mahan1967}, where the dielectric constant for DC fields, $\epsilon(0)$, is equal to $11$ for the case of yttrium orthosilicate (Y$_2$SiO$_5$) \cite{Christiansson2001, Mock2018}. $\varepsilon_0$ is the vacuum permittivity and $h$ is Planck's constant. Implicit in this equation is the reasonable assumption that the static dipole moment difference remains the same regardless of the states of other ions.

\section{\label{app:host_crystal}Host crystal and unit cell}

\begin{figure*}
\includegraphics[width=\textwidth]{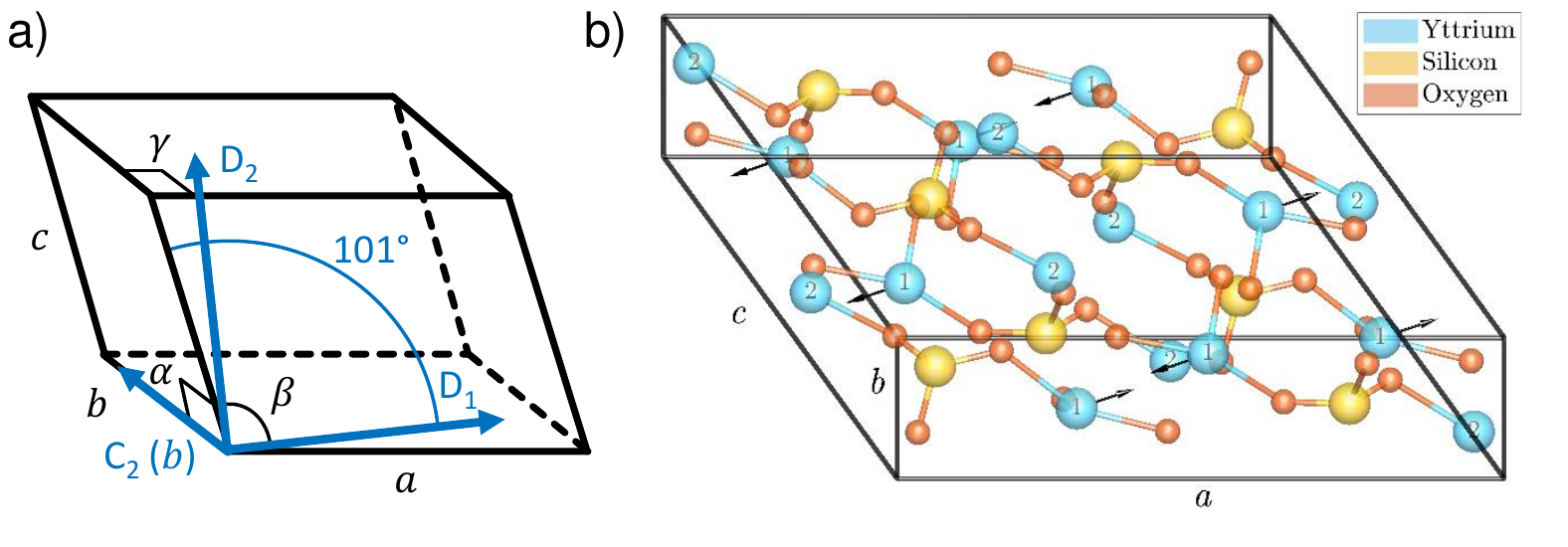}
\caption{\label{fig:Crystal}a) Shows the unit cell of Y$_2$SiO$_5$ in the C2/c (base-centered) space group, see Tab. \ref{tab:YSO} for values of the parameters. The figure also indicates the connection to the principal axes D$_1$, C$_2$ which is more often denoted by b, and D$_2$. b) Shows the layout of the $8$ basic Y$_2$SiO$_5$ molecules that each unit cell contains constructed using the atom coordinates of Ref. \cite{Villars2021} and the symmetries of the C2/c space group. Note that the connections shown are only there to indicate which basic molecule the ions belong to. Two yttrium crystal sites exist with coordination numbers of $6$ and $7$, which we label as site $1$ and $2$, respectively. We assume that the static electric dipole moment difference, $|\Delta\boldsymbol{\mu}| = 7.6 \cdot 10^{-32}$ Cm \cite{Graf1998}, points in the D$_1$ direction for the site 1 yttrium ion belonging to the bottom left basic molecule. Since the basic molecules are oriented differently, we indicate the direction of $\Delta\boldsymbol{\mu}$ by black arrows for all site 1 yttrium ions.}
\end{figure*}

In order to understand how the positions and orientations of dopants are obtained, this section discusses the structure of the host crystal, whose unit cell is described in Fig. \ref{fig:Crystal}a and Tab. \ref{tab:YSO}, where the relation to the principal axes $D_1$, $b$, and $D_2$ is also indicated. Each unit cell contains $8$ basic molecules of Y$_2$SiO$_5$ as seen in Fig. \ref{fig:Crystal}b. The Eu dopant can replace either of the two Y ions in the basic molecule of Y$_2$SiO$_5$, denoted by the numbers $1$ and $2$ in the figure. These crystal sites differ in that the Y ions in $1$ have $6$ nearby oxygen, whereas ions in $2$ have $7$ (coordination numbers of $6$ and $7$, respectively). Note that these crystal sites are the same as the sites normally referred to when discussing spectral properties, e.g., optical transition wavelengths, hyperfine energy level splittings, etc. However, to the best of the authors knowledge there hasn't been a clear demonstration of which crystal site ($6$ or $7$ nearby oxygen) correspond to which spectral site for the high temperature phase X2 crystal structure of Y$_2$SiO$_5$, and the definition varies between Refs. \cite{Lin1996, Malyukin2002, Wen2014, Mirzai2021}. We here assume that the spectral site denoted by $1$ with the properties shown in Fig. \ref{fig:Eu_Enery_levels}a-b has a coordination number of $6$. This is assumed since spectral site $1$ has larger crystal field splittings \cite{Konz2003} which indicates that it corresponds to coordination number $6$ where the distances between the dopant and the nearby oxygens are shorter, following the same reasoning as in Ref. \cite{Welinski2019}. Note that even if this assumption is wrong the general results of Sec. \ref{sec:ISD_crystal_analysis} are still valid since using the crystal site with coordination number $7$ generates very similar results. 

The $8$ basic molecules in each unit cell are oriented in four different ways resulting from the crystal symmetries of identity, inversion, mirror of $b$, and the combination of inversion and mirror of $b$. These orientations are indicated by the black arrows of Fig. \ref{fig:Crystal}b, which show the assumed direction of the static electric dipole moment difference $\Delta\boldsymbol{\mu}$ used to calculated the dipole-dipole shift from Eq. \ref{eq:delta_nu}. The magnitude $|\Delta\boldsymbol{\mu}| = 7.6 \cdot 10^{-32}$ Cm is known \cite{Graf1998}, but to the best of the authors knowledge the direction in relation to the basic molecule is not. We here assume that the direction is along the $D_1$ principal axis for the basic molecule shown in the bottom left corner of Fig. \ref{fig:Crystal}b. Note that this choice was an arbitrary one. However, once more the general results of Sec. \ref{sec:ISD_crystal_analysis} are still valid even if this assumed direction is wrong. 

\begin{table}[b]
\caption{\label{tab:YSO}Unit cell parameters for the high temperature phase X2 crystal structure of Y$_2$SiO$_5$, see Fig. \ref{fig:Crystal}a,  written in the C2/c (base-centered) space group \cite{Maksimov1971,Bengtsson2012}. Note that the parameters are sometimes given in the I2/c (body-centered) space group instead.}
\begin{ruledtabular}
\begin{tabular}{ll}
\textrm{Distances}&
\textrm{Angles}\\
\colrule
$a = 1.44137$ nm & $\alpha = 90^\circ$\\
$b = 0.6719$ nm & $\beta = 122.235^\circ$\\
$c = 1.040$ nm & $\gamma = 90^\circ$\\
\end{tabular}
\end{ruledtabular}
\end{table}

A qubit surrounding is created by first randomly replacing a fraction of the site $1$ Y ions with $^{153}$Eu ions. In experiments the total doping concentration, $c_\text{total}$, of replacing the Y ions in either site $1$ or site $2$ is often cited, but to the best of the authors knowledge the exact relative occupation of these two sites is unknown. There are indications that the substitution favors the site with higher coordination number \cite{Wen2014, Mirzai2021}, i.e., site $2$ using our definition of crystal sites. However, this might be more important for dopants such as praseodymium or cerium which are larger than the yttrium ion they replace \cite{Serrano2014, Malyukin2002, Wen2014}. For europium, whose size is more comparable with yttrium, the site occupation may be more equal \cite{Serrano2014, Konz2003}. Regardless, since the exact relative occupation is still unknown we here assume that they are equal, i.e., half of the total number of dopants are in site $1$ and half are in site $2$. To be consisted with the experimental concentration values often quoted in articles, the concentration values used throughout this work refers to the total number of $^{153}$Eu ions in the crystal, but only half of those are assumed to be in site $1$ and those are the only ions that are used to evaluate ISD. After all ions have been placed their positions in space is known, and the dipole-dipole shift between any two ions can be calculated using Eq. \ref{eq:delta_nu}. 

\section{\label{app:interpolation_methods}Estimating ISD from non-qubit ions}
To use the QBies method described in Eq. \ref{eq:extrapolation}, the rotation and shrinkage of the qubit $0$ Bloch vector that each non-qubit ion causes must first be determined. However, there are millions of ions to investigate, and even though the fast QBies method is used it would take prohibitively long time to simulate everything if a new simulation is done for each non-qubit ion. Fortunately, one can use interpolation to heavily reduce the computational time. How this interpolation is performed differs for non-qubit ions inside or outside the reserved frequency range of qubit $0$, and is explained further in the following subsections. 

\subsection{\label{app:interpolation_inside}Non-qubit ions inside the reserved frequency range of qubit $0$}
\begin{figure}
\includegraphics[width=\columnwidth]{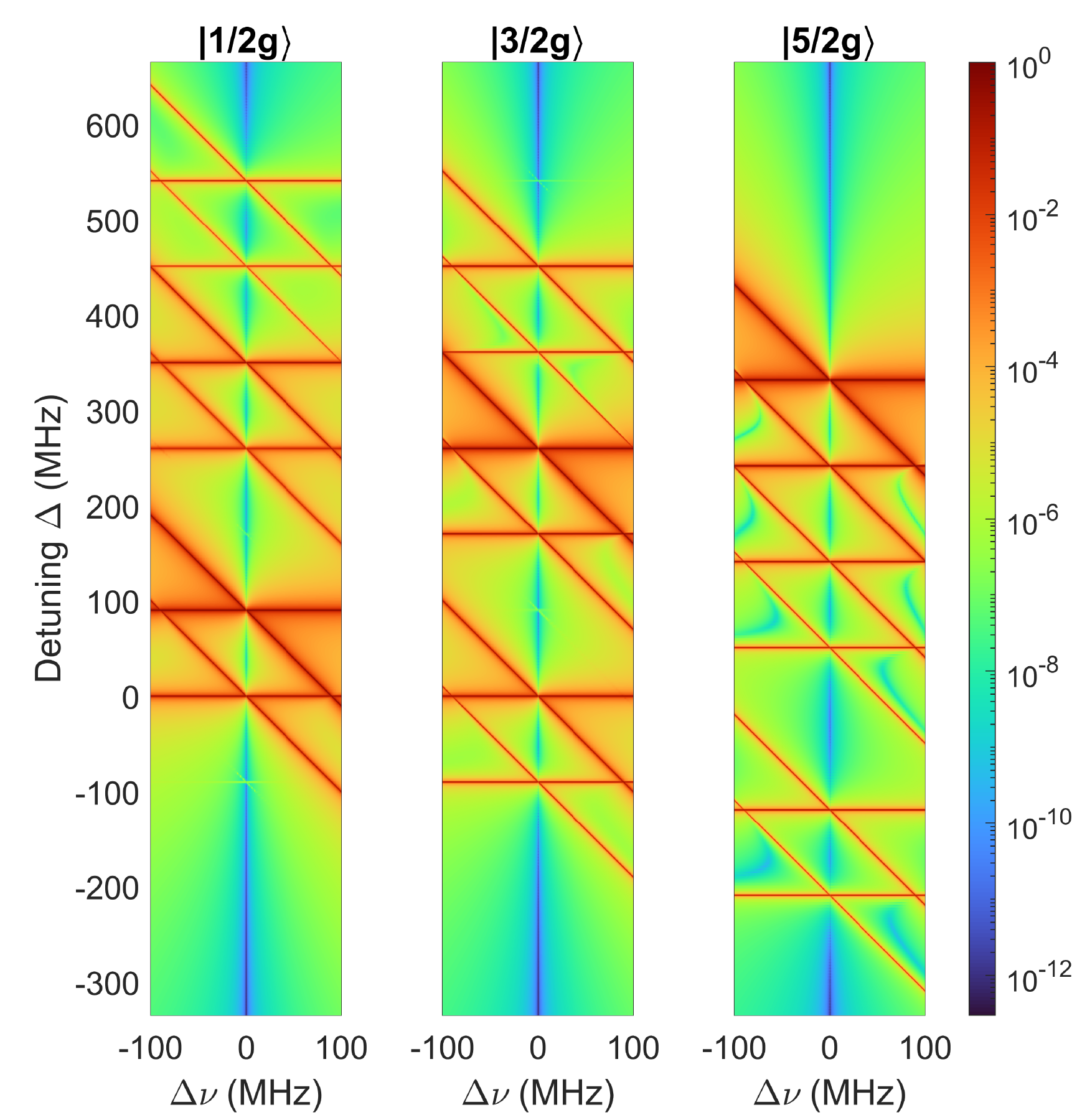}
\caption{\label{fig:Shift_detuning_map}Shows the additional SQ gate error for a qubit due to ISD from one non-qubit ion. The dipole-dipole shift, $\Delta\nu$, and the detuning, $\Delta$, of the non-qubit ion are varied on the horizontal and vertical axes, respectively, and the three graphs show the situation for different initial states of the non-qubit ion: $|1/2g\rangle$, $|3/2g\rangle$, and $|5/2g\rangle$, respectively. In all cases the qubit starts in $|0\rangle+i|1\rangle$ and a NOT operation is attempted.}
\end{figure}

In order calculate how ISD affects the SQ gate error in a realistic case, the pulses intended to drive the $|0\rangle\rightarrow|e\rangle$ and $|1\rangle\rightarrow|e\rangle$ transitions of the qubit are now allowed to drive any of the nine optical transitions in the non-qubit ion, i.e., all six energy levels of the non-qubit ion are now included in contrast to how it was treated in the idealized case studied in Sec. \ref{sec:general_ISD}. Note however, that the pulses still only drive the intended transitions in the qubit to be able to separate the errors due to internal crosstalk from ISD. The fact that the non-qubit ion now has nine optical transitions complicates how ISD depend on the dipole-dipole shift and detuning as can be seen in Fig. \ref{fig:Shift_detuning_map} where the ISD error is seen for the case of one non-qubit ion. Here the non-qubit ion starts in one of the three ground states $|1/2g\rangle$, $|3/2g\rangle$, or $|5/2g\rangle$, has a detuning of $\Delta$ relative to the qubit, and interacts with a dipole-dipole strength of $\Delta\nu$. As before the qubit starts in $|0\rangle+i|1\rangle$ and a NOT operation is attempted. In each of the three graphs six horizontal lines of high errors can be seen. These correspond to detunings where the pulses driving $|0\rangle\rightarrow|e\rangle$ or $|1\rangle\rightarrow|e\rangle$ in the qubit are also resonant with an available transition in the non-qubit ion, i.e., a transition from the the starting ground state of the non-qubit ion to any of the three excited states. The differences in thickness between these lines come from differences in the relative oscillator strengths of the transitions being driven and the intended transitions. Furthermore, the six horizontal lines in the $|3/2g\rangle$ case are shifted by $-90$ MHz compared to the lines in the $|1/2g\rangle$ case since the splitting between the two hyperfine levels is $90$ MHz. When the non-qubit ion is far detuned from the optical pulses, the effect of ISD is low, except when the dipole-dipole shift is such that it compensates for the detuning, which can be seen in the diagonal lines showing high errors. In these cases the error is large since the initially detuned non-qubit ion is shifted into resonance through the dipole-dipole interaction when the qubit ion is excited during the gate operation. The combinations of dipole-dipole shifts and detunings that cause a significant additional error due to ISD are in the vicinity of these horizontal and diagonal lines. 

For non-qubit ions inside the $-335$ MHz to $665$ MHz reserved frequency range of qubit $0$, we estimate the qubit $0$ Bloch vector using a bilinear interpolation of the data underlying the results of Fig. \ref{fig:Shift_detuning_map}. After the Bloch vector is determined the rotation and shrinkage compared to the targeted state can easily be calculated. This interpolation is used for all non-qubit ions within the reserved frequency range listed above except for those few cases where the non-qubit ion had a dipole-dipole shift magnitude larger than $100$ MHz. Note that it is unlikely that ions cause larger dipole-dipole shifts than $\pm100$ MHz, e.g., in several thousand different simulations where each simulation, depending on concentration, contains between $500$ and $7000$ ions within the reserved frequency range of qubit $0$, only $13$ ions had a dipole-dipole shift magnitude larger than $100$ MHz. For the ions with large shifts, new simulations with their specific detunings and shifts were performed to obtain the resulting Bloch vector of qubit $0$.

The reserved frequency range of $-335$ MHz to $665$ MHz was picked for two different reasons. First, it covers all the diagonal lines of high errors shown in Fig. \ref{fig:Shift_detuning_map} for dipole-dipole shifts of $\pm 100$ MHz with roughly $25$ MHz to spare. Where, again, it is relatively unlikely for an ion to cause a shift greater than $100$ MHz. Second, the total reserved frequency range is $1$ GHz, which is equal to the separation between different qubits, thus making it easy to assign each non-qubit ion to the corresponding qubit whose pulses can affect them, as is shown in Fig. \ref{fig:Eu_Enery_levels}d. 

\begin{figure*}
\includegraphics[width=\textwidth]{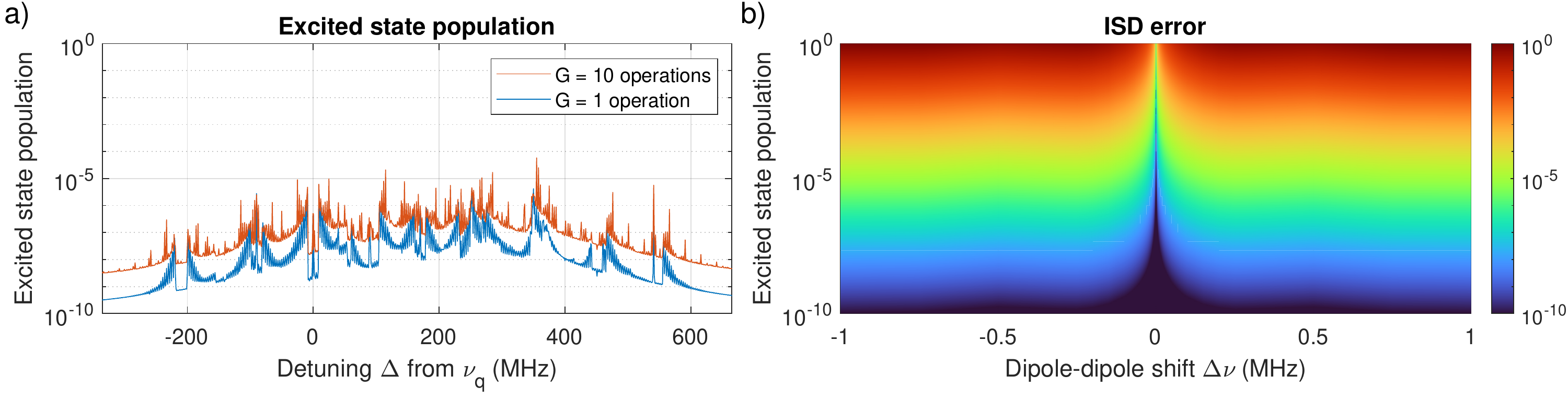}
\caption{\label{fig:ISD_excitation}A non-qubit ion is assigned a corresponding qubit based on its detuning from qubit $0$ as shown in Fig. \ref{fig:Eu_Enery_levels}d. The non-qubit ion can be off-resonantly excited by the pulses intended to perform $G$ NOT operations on its corresponding qubit. a) Shows the total population in the excited states as a function of the detuning of the non-qubit ion to its corresponding qubit, whose frequency of its $|0\rangle\rightarrow|e\rangle$ transition is $\nu_\text{q}$. The detuning of the non-qubit ion always lies in the range of $-335$ MHz to $665$ MHz, since it would otherwise be assigned to a different corresponding qubit. The complex dependence on the detuning comes from the fact that the initial state of the non-qubit ions depend on their detuning due to the creation of transmission windows surrounding the two qubit transitions in the same way as described in Sec. \ref{sec:ISD_Crystal} and Appendix \ref{app:transmission_windows}. b) A non-qubit ion causes an additional SQ gate error on qubit $0$ due to ISD if it is partly excited when the gate operation on qubit $0$ is performed. This additional error is shown in the graph as a function of the dipole-dipole shift $\Delta\nu$ (horizontal axis) between the non-qubit ion and qubit $0$, and as a function of the total excited state population (vertical axis) of the non-qubit ion before the gate operation on qubit $0$ is attempted.}
\end{figure*}

\subsection{\label{app:interpolation_outside}Non-qubit ions outside the reserved frequency range of qubit $0$}
Non-qubit ions outside the reserved frequency range of qubit $0$ are assumed to not interact with the pulses intended to drive qubit $0$. However, such ions can still cause ISD to qubit $0$ if they are partly excited before the gate operation on qubit $0$ is attempted. This section describes how the contribution of ISD from such non-qubit ions is estimated. 

First, we determine which qubit index each non-qubit ion belong to, e.g., an ion with a detuning of $2.4$ GHz from qubit $0$ belong to qubit index $3$ as shown in Fig. \ref{fig:Eu_Enery_levels}d. Then the detuning between the non-qubit ion and its corresponding qubit is used to determine the probabilities of starting in the three different ground states, for more information see Appendix \ref{app:transmission_windows}.

Second, we estimate how large fraction of the non-qubit ion population is in the excited state after $G$ gate operations have been performed on its corresponding qubit. This is done using a linear interpolation based on simulations where the non-qubit ions had detunings between $-335$ MHz to $665$ MHz. We only keep the information of how large fraction of the population there is in total in any of the three excited states, since this is the only factor which impacts the effect of ISD. This total population in the excited states as a function of detuning is shown in Fig. \ref{fig:ISD_excitation}a for the cases when $1$ or $10$ gate operations were performed. These simulations included the effects of decay, decoherence, and internal crosstalk to also model the slight decay which occur when performing multiple gate operations. 

Third, new simulations are performed to estimate the ISD error on qubit $0$ due to dipole-dipole interaction with non-qubit ions that are initially partly excited, but does not interact with any pulses during the gate operation performed on qubit $0$. In these simulations the effect of different dipole-dipole shifts is studied as normal, but the amount of excitation of the non-qubit ion, which now only has one ground and one excited state, is also varied. To be able to isolate the ISD errors these simulations did not include decay, decoherence, and internal crosstalk. The results from these simulations can be seen in Fig. \ref{fig:ISD_excitation}b. As can be seen, the error is constant for large magnitudes of the dipole-dipole shift. Simulations were performed up to $\pm100$ MHz and any non-qubit ion causing a shift larger than this was assumed to affect the qubit Bloch vector in the same way as an ion with a $100$ MHz shift. 

In summary, the detuning of the non-qubit ion from its corresponding qubit gives a probability distribution to be in the three ground states, which together with the detuning determines how large fraction of the population is excited after $G$ gate operations have been performed on the corresponding qubit, see Fig. \ref{fig:ISD_excitation}a. The fraction of the population which is in the excited states is then used to estimate the qubit $0$ Bloch vector due to ISD from the non-qubit ion through a linear interpolation of the simulation data underlying the results shown in Fig. \ref{fig:ISD_excitation}b. Finally, the rotation and shrinkage of the qubit $0$ Bloch vector when compared to the targeted state is calculated.

\section{\label{app:transmission_windows}Creating transmission windows}
In this section we describe the procedure to simulate the creation of the transmission windows seen in Fig. \ref{fig:Eu_Enery_levels}e using spectral hole burning techniques. First, one can calculate the largest possible widths of such transmission windows by iterating through all inhomogeneously broadened ions and placing them in the ground state whose transitions to all excited states are as far away as possible from the frequencies of the two qubit transitions \cite{Kinos2021a}. For site 1 $^{153}$Eu:Y$_2$SiO$_5$ the empty regions surrounding the two optical transitions, $|0\rangle \rightarrow |e\rangle$ and $|1\rangle \rightarrow |e\rangle$, range from $-9.0$ MHz to $9.1$ MHz, and $-35.9$ MHz to $14.6$ MHz, measured from the center of the respective transitions. Based on this knowledge we send in frequency scanning pulses to empty these regions from any absorbing ions. However, since we assume that our pulses have some widths and can also off-resonantly excite ions we make the frequency scanning regions roughly $1$ MHz narrower compared to the values listed above. 

\begin{figure}
\includegraphics[width=\columnwidth]{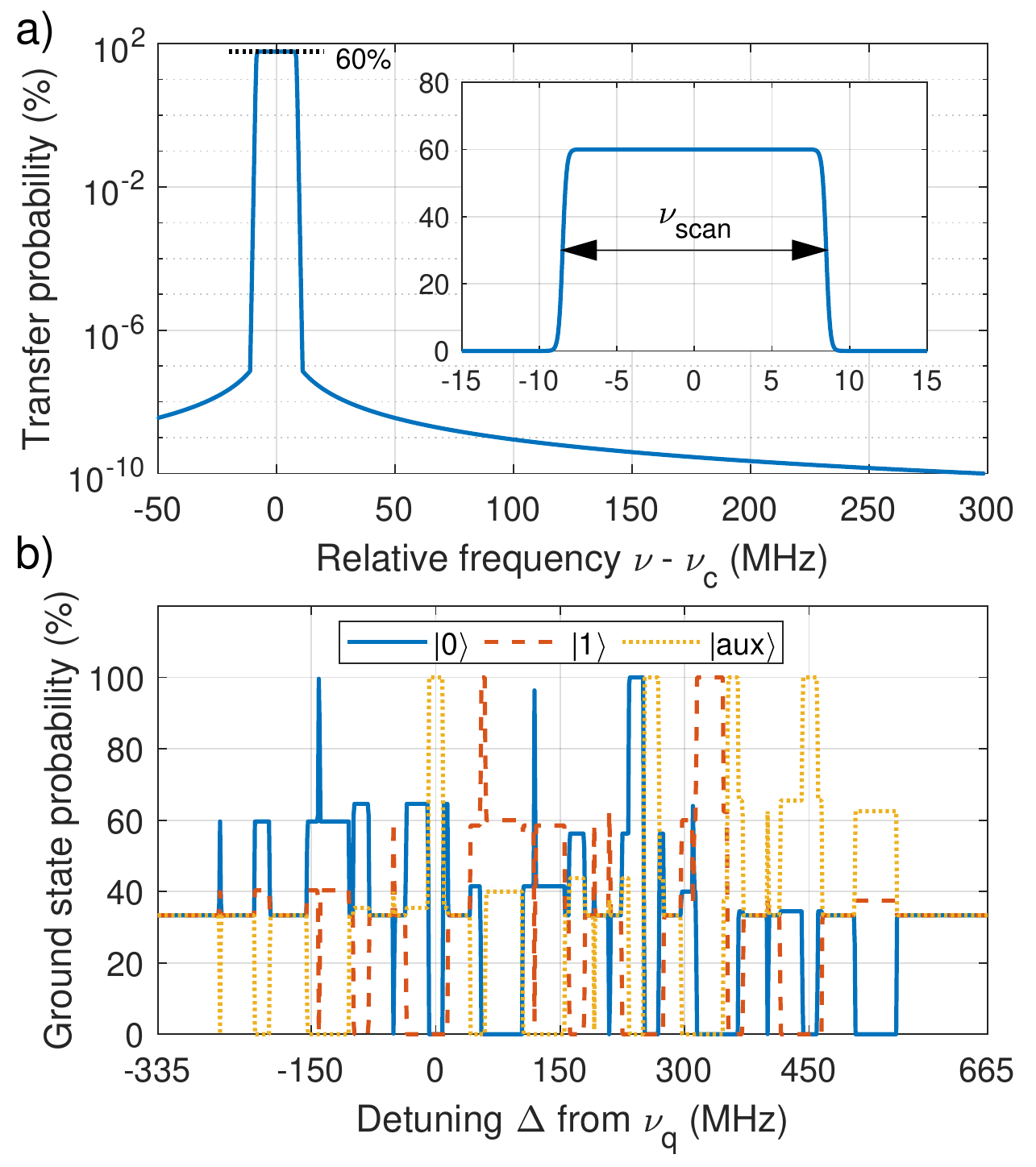}
\caption{\label{fig:Holeburning}a) Shows the probability to transfer an ion from one ground state to one excited state as a function of the relative frequency of that transition, $\nu$, when compared to the center frequency of the incoming light pulse, $\nu_c$. Here we show the example of pulse $\#1$ in Tab. \ref{tab:pulses} for the $|0\rangle \rightarrow |5/2e\rangle$ transition. The probability is calculated using Eq. \ref{eq:HoleBurning}. The transfer is most efficient within a frequency range of $\nu_\text{scan}$ as is show in the inset, which shows the same transfer probability but in linear scale. The maximum transfer efficiency (here shown as $60\%$) depends on which transition is being driven which is explained further in Tab. \ref{tab:pulses}. b) Shows the probability to be in each of the three ground states as a function of the detuning $\Delta$ from the $|0\rangle \rightarrow |e\rangle$ transition frequency, $\nu_q$, of qubit $q$. Here $\nu_\text{init} = 0$ kHz is used. These ground state population probabilities results in the transmission windows seen in Fig. \ref{fig:Eu_Enery_levels}e.}
\end{figure}

Performing such hole burning simulations using the Lindblad master equation would take prohibitively long time since it needs to keep track of several hundreds of thousands of ions for several thousand incoming light pulses. Therefore, these simulations are performed in a simplified way where we do not keep track of any coherence and instead assume a specific transfer efficiency for each pulse that is applied. The simulations keep track of the population in all six energy levels as a function of the detuning from the qubit, ranging from $-335$ MHz to $665$ MHz. Any light pulses that are sent in have a probability as a function of frequency to transfer the population to an excited state as seen in Fig. \ref{fig:Holeburning}a. This probability is calculated in the following way: 

\begin{eqnarray}\label{eq:HoleBurning}
  P_1(\nu) &&= \frac{\tanh\left(\frac{\nu - (\nu_c - \frac{\nu_\text{scan}}{2})}{\nu_\text{slope}}\right) - \tanh\left(\frac{\nu - (\nu_c + \frac{\nu_\text{scan}}{2})}{\nu_\text{slope}}\right)}{2}\nonumber\\
  P_2(\nu) &&= \frac{\gamma^2}{(\nu - \nu_c)^2 - \gamma^2} \nonumber\\
  P(\nu) &&= \epsilon_i \cdot \text{max}(P_1(\nu), P_2(\nu))
\end{eqnarray}

where $\nu$ is the frequency of the transition being examined for a certain ion, and $\nu = 0$ MHz for the $|0\rangle \rightarrow |e\rangle$ transition frequency of the qubit ion. Furthermore, $\nu_c$ and $\nu_\text{scan}$ are the center and scanning frequencies of an incoming light pulse. These determine the frequency region within which the transfer is efficient as can be seen in the inset of Fig. \ref{fig:Holeburning}a. $\nu_\text{slope}$ determines the slope of the transfer probability outside the $\nu_\text{scan}$ frequency range. $P_1(\nu)$ models the transfer distribution for frequencies close to the light frequency, but in order to also model the off-resonant excitation, which might occur for ions that are heavily detuned from the light pulses, we assume a Lorentzian fall off which is specified in $P_2(\nu)$ where $\gamma = 1/T_2$ and $T_2 = 2.6$ ms. Lastly, we assume that the probability to transfer a specific ion on a specific transition $i$ is given by $P(\nu)$ where $\epsilon_i$ is the scaled transfer efficiency for that specific transition, which is explained further in Tab. \ref{tab:pulses}.

\begin{table*}
\caption{\label{tab:pulses}Specification of the pulses used to prepare the transmission windows. Each pulse scans a frequency region of $\nu_c \pm \nu_\text{scan}/2$, where $\nu_c$ is the center frequency which is given relative to the frequency of the $|0\rangle \rightarrow |e\rangle$ transition of the qubit, and $\nu_\text{scan}$ is the total frequency range. The qubit initialization pulses, $\#4$ and $\#5$, are examined for a few different frequency widths: $\nu_\text{init} = 100$ kHz, $75$ kHz, $50$ kHz, $25$ kHz, and $0$ kHz. When $\nu_\text{init} = 0$ kHz no initialization pulses ($\#4$ and $\#5$) are used and only the qubit ion is transferred. $\nu_\text{slope}$ sets the slope of the transfer probability distribution outside the $\nu_\text{scan}$ range as seen in the inset of Fig. \ref{fig:Holeburning}a. The last two columns specifies what transfer efficiency the pulse have for a specific transition. The efficiency of transferring population along another transition, $\epsilon_i$, is scaled linearly with the ratio of the square root of the relative oscillator strengths for the new transition relative the designed transition, where the maximum transfer efficiency is capped at $100\%$. In Tab. \ref{tab:scheme} the hole burning pulse sequence is shown.}
\begin{ruledtabular}
\begin{tabular}{llllll}
\textrm{Pulse $\#$}&\textrm{Name}&\textrm{$\nu_c \pm \nu_\text{scan}/2$ (MHz)}&\textrm{$\nu_\text{slope}$ (kHz)}&\textrm{Design transition}&\textrm{Efficiency for transition}\\
\colrule
1 & Window $|0\rangle$ & $0 \pm 17/2$ & $250$ & $|0\rangle \rightarrow |5/2e\rangle$ & $60\%$ \\
2 & Window $|1\rangle$ & $79.35 \pm 49.3/2$ & $250$ & $|0\rangle \rightarrow |5/2e\rangle$ & $60\%$ \\
3 & Clear $|\text{aux}\rangle$ & $-50.8 \pm 1/2$ & $250$ & $|\text{aux}\rangle \rightarrow |3/2e\rangle$ & $60\%$ \\
4 & Qubit excite & $-50.8 \pm \nu_\text{init}/2$ & $\nu_\text{init}/4$ & $|\text{aux}\rangle \rightarrow |3/2e\rangle$ & $99.9\%$ \\
5 & Qubit deexcite & $-260 \pm 10\cdot \nu_\text{init}/2$ & $10\cdot \nu_\text{init}/4$ & $|0\rangle \rightarrow |3/2e\rangle$ & $99.9\%$ \\
\end{tabular}
\end{ruledtabular}
\end{table*}

\begin{table*}
\caption{\label{tab:scheme}Shows the pulse sequence used to create transmission windows. Information about the various pulses can be seen in Tab. \ref{tab:pulses}. If decay is included any population in the excited states after a pulse is completed is fully transferred to the ground states using branching ratios that are equal to the relative oscillator strengths shown in Fig. \ref{fig:Eu_Enery_levels}b. If decay is not included the population in the excited state remains and can thus be transferred back to the ground state via subsequent pulses. The sequence is applied in the following way: For each goal, the first pulse is applied to all qubits, then if decay is included any excited state population decays, before the second pulse is applied to all qubits. This continues until all pulses have been applied, and is then repeated a certain number of times before moving onto the next goal. After all five goals have been completed the transmission windows for all qubits have been prepared. }
\begin{ruledtabular}
\begin{tabular}{llll}
\textrm{Goal}&\textrm{Pulses ($\#$)}&\textrm{Repetitions of each pulse}&\textrm{Include decay}\\
\colrule
Clear windows and $|\text{aux}\rangle$ & 1, 2, 3 & 20 & true \\
Clear windows & 1, 2 & 500 & true \\
Excite qubit & 4 & 1 & false \\
Deexcite qubit & 5 & 1 & false \\
Clear second window & 2 & 100 & true \\
\end{tabular}
\end{ruledtabular}
\end{table*}

All pulses used are defined in Tab. \ref{tab:pulses}, and the order in which they are applied is described in Tab. \ref{tab:scheme}. The transmission windows are created in the following way. The frequency regions surrounding the two qubit transitions, $|0\rangle \rightarrow |e\rangle$ and $|1\rangle \rightarrow |e\rangle$, are emptied of almost all absorbing ions. This is done in two steps, where the first also tries to empty ions with frequencies close to the qubit $|\text{aux}\rangle \rightarrow |3/2e\rangle$ transition. Since it is impossible to remove ions from all ground states simultaneously, this first step only removes ions that are far detuned from the qubit ion. The second step only cleans the regions around the two qubit transitions, which is possible for all ions, thus creating two transmission windows with very little absorption. After these steps the qubit ion is in the $|\text{aux}\rangle$ state and must be transferred back into $|0\rangle$. This is done by first exciting on $|\text{aux}\rangle \rightarrow |3/2e\rangle$ and then deexciting on $|0\rangle \rightarrow |3/2e\rangle$. These initialization pulses can create some residual absorption in the second transmission window surrounding $|1\rangle \rightarrow |e\rangle$, and the last step therefore cleans this second window. 

After creating the transmission windows of qubit $0$ the residual absorption remaining in the windows depend on how many other qubits have been prepared. We assume that the qubits are separated by $1$ GHz and number them as shown in Fig. \ref{fig:Eu_Enery_levels}d. If we prepare each qubit after one another, i.e., prepare qubit $0$ first, then qubit $1$ etc., the residual absorption in the transmission windows of the first qubit is growing to unacceptable levels. However, one can instead apply the hole burning pulses in an interleaved way, i.e., the first burning pulse in the sequence described in Tab. \ref{tab:scheme} is applied for all qubits before moving onto the second burning pulse in the sequence. If this interleaved burning is used, then the residual absorption saturates at an negligible level after preparing a few tens of qubits. In our final simulations we send in light pulses to prepare transmission windows for $51$ qubits, but only keep the population distribution for the central qubit, i.e., qubit $0$. Since transmission windows are created for each qubit and we use the same population distribution for all cases, the probabilities as a function of detuning is periodic with a period of $1$ GHz. An example of this population distribution can be seen in Fig. \ref{fig:Holeburning}b, where the frequency width of the initialization pulses are $\nu_\text{init} = 0$ kHz. This population distribution is used to determine the probabilities of a non-qubit ion starting in each of the three ground states as a function of detuning from its corresponding qubit ion.

\section{\label{app:ISD_N_and_r}ISD dependence on the number of ions contributing and their distances from the qubit}
\begin{figure}
\includegraphics[width=\columnwidth]{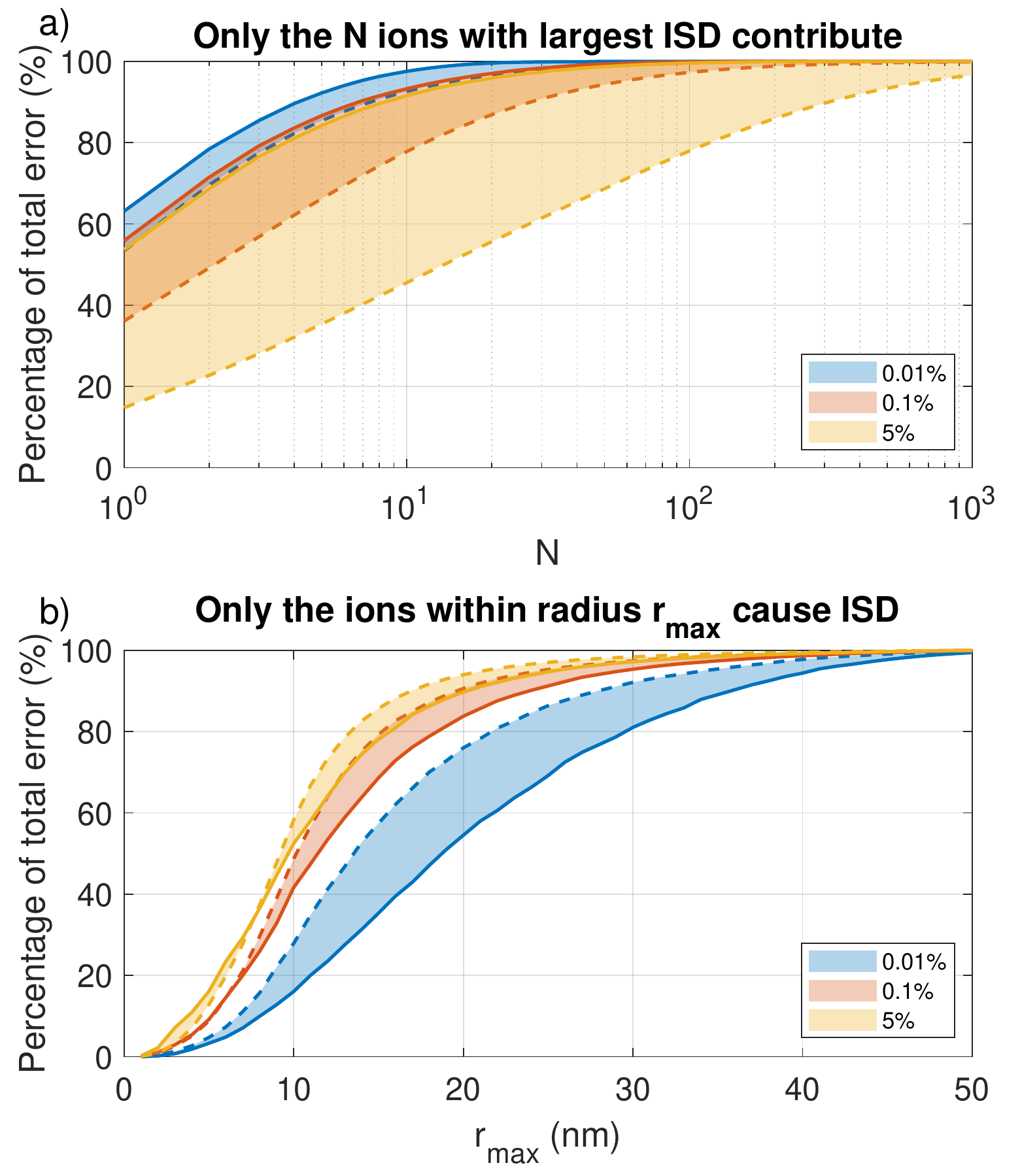}
\caption{\label{fig:ISD_N_and_r}Shows the percentage of the total ISD error that on average is accounted for when a) only the N ions with the largest ISD and b) only the ions within radius $r_\text{max}$ contribute to the error. Colors indicate different total doping concentrations. Solid (dashed) lines show the results when $0$ ($10$) gate operations were performed on each of the qubits labeled $1\rightarrow 50$ before attempting the NOT operation on qubit $0$.}
\end{figure}

This section studies how large the ISD errors are if only the $N$ ions with the largest individual ISD errors are considered, or if only the ions that spatially lie within a radius $r_{max}$ from the qubit are considered. The results of these investigations can be seen in Fig. \ref{fig:ISD_N_and_r}. In general, more ions contribute to the total error when the concentration is high or when the number of previous gate operations are high. In contrast, the spatial radius within which the ions that cause the largest fraction of the ISD error lies, is shorter when the concentration or the number of previous gate operations are high. The results in Fig. \ref{fig:ISD_N_and_r} can be used to estimate the qubit surrounding size and how many ions need to be included in future simulations to give a good estimate of the ISD error.

\section{\label{app:ISD_detailed_analysis}Detailed analysis of the ISD error}
This section provides a detailed analysis of Fig. \ref{fig:ISD_q_and_op}. A few things should be noted. 

First, for low doping concentrations the ISD error when running $G$ gate operations on $10$ or $50$ additional qubits is roughly the same. This is because for low doping concentrations the inhomogeneous absorption profile is relatively narrow, see Eq. \ref{eq:Gamma_inh}. Therefore, qubits with large indices $q$ which are heavily detuned from the center of the inhomogeneous absorption profile, see Fig. \ref{fig:Eu_Enery_levels}d, do not have many ions within their reserved frequency range. Fewer ions are thus partly excited and hence no significant ISD error is added due to running gate operations on those qubits. However, as the concentration increases the number of ions belonging to qubits with high $q$-indices grows, and thus the additional ISD errors also grow.

Second, some curves showing the ISD error as a function of the ordered simulation number have steeper slopes than others, meaning that the ISD error can vary drastically depending on the exact surrounding of the qubit. This occurs when the total number of ions involved in causing ISD is relatively low, e.g., when the doping concentration is low or when $Q$ is low. This is true even if $G$ is large, since $G$ does not change how many ions contribute. However, when the doping concentration is high and $Q$ is large, all simulations yield roughly the same additional error, see, e.g., the purple data of Fig. \ref{fig:ISD_q_and_op}f. This is reasonable since in this case there are many ions that can potentially cause ISD, thus averaging out the statistical likelihood that a certain error occurs. 

\begin{figure}
\includegraphics[width=\columnwidth]{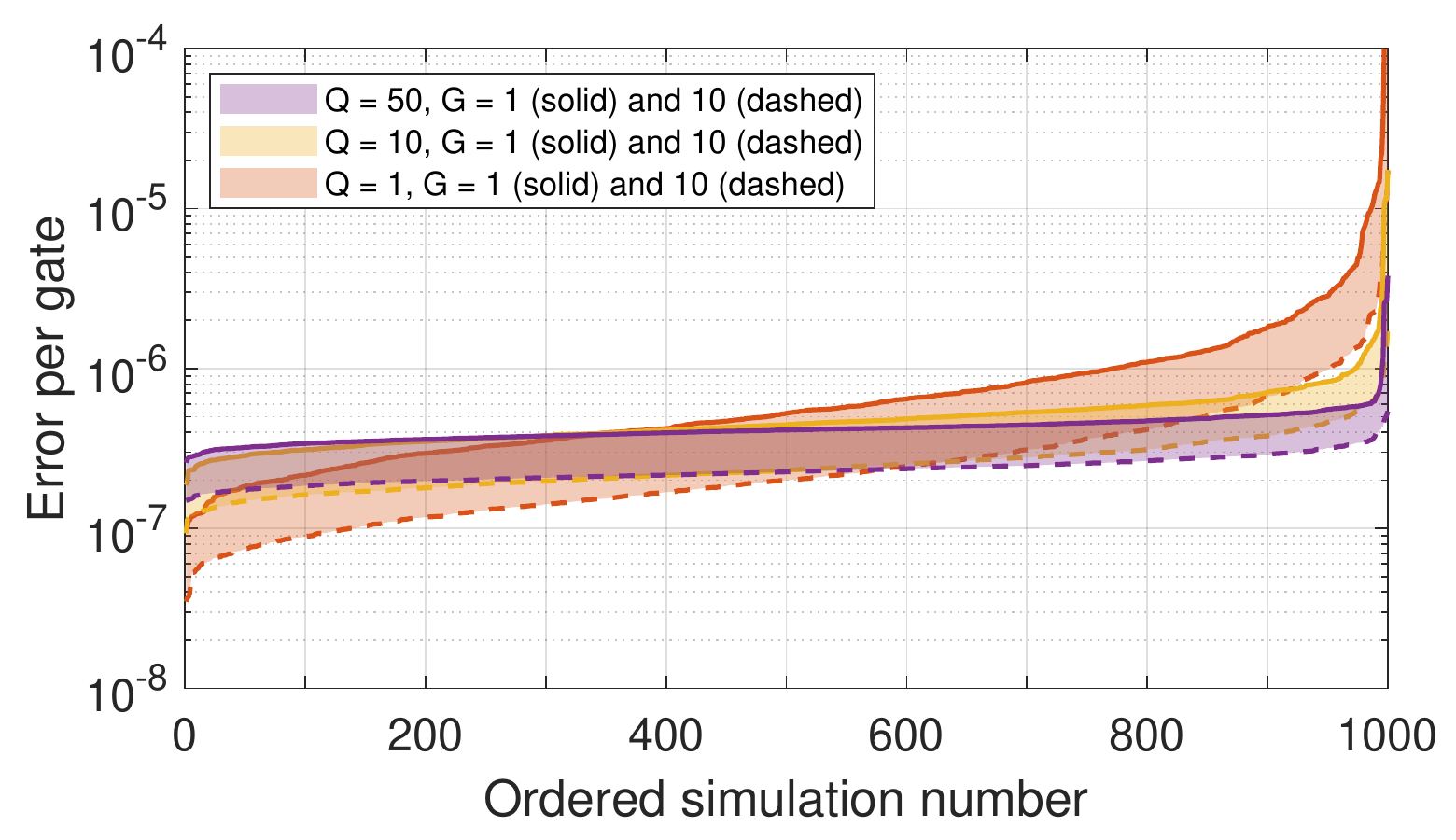}
\caption{\label{fig:ISD_q_and_op_per_gate}Shows the ISD error divided by $G\cdot Q$, i.e., how much error is on average added due to performing a single gate operation on one other qubit before the gate operation on qubit $0$ is attempted. The total doping concentration is $c_\text{total} = 5\%$. }
\end{figure}

A third observation is made when studying the higher concentrations where all qubits have roughly the same number of ions within their reserved frequency range. The observation is that the additional error due to ISD is increased by roughly $4\cdot10^{-7}$ for each gate operation that is performed on another qubit before the gate operation on qubit $0$ is attempted, as can be seen in Fig. \ref{fig:ISD_q_and_op_per_gate}. In other words, for high concentrations the ISD error scales linearly with the total number of gate operations performed before the gate operation on qubit $0$ is attempted. 

This linear scaling is in our model reasonable since more gate operations lead to more excitation which in turn lead to higher ISD errors. However, in a realistic situation the ions would eventually decay from the excited state due to the limited lifetime, and therefore no longer cause an additional ISD error. This is not fully included in our model where the limited life- and coherence times of the ions are only included during the up to ten gate operations applied on its corresponding qubit. In other words, it is as if the ten gate operations on the up to $50$ qubits are applied in parallel before the gate operation on qubit $0$ is attempted. In reality such gate operations would most likely have to be performed sequentially to prevent unwanted dipole-dipole interactions between different qubits. 

Fortunately, one can estimate when the decay of ions starts to affect the results shown in Fig. \ref{fig:ISD_q_and_op} based on the SQ gate duration of $3.36$ $\upmu$s and the assumed optical lifetime of $1.9$ ms \cite{Equall1994}. With no downtime between gate operations the total duration to perform $G$ gate operations on $Q$ qubits is $G\cdot Q\cdot 3.36$ $\upmu$s. Fig. \ref{fig:ISD_q_and_op} do not study ISD for more than $10$ operations on $50$ different qubits because the total duration needed to apply those gates is $1.68$ ms, which means that already in this case the decay of ions would probably start to affect the results. 

At first glance one might therefore expect a saturation in the fraction of excited non-qubit ions, and thus a saturation in the ISD error due to previous gate operations. Unfortunately, an ion might not return to its original ground state after decaying. In the worst case, an ion originally absorbing outside the transmission windows could after being excited decay into another ground state which has a transition frequency inside the transmission windows. Thus heavily increasing the risk that the ion is excited once more. Despite this, it is reasonable to assume that the linear growth of the ISD error slows down somewhat when more gate operations are performed, since there is still a chance that the excited ions decay back into their original ground states. However, we note that more simulations are needed to correctly estimate the long-term ISD error in a rare-earth quantum computer when running even more gate operations.

A fourth and final note can be made when one considers that the inhomogeneous absorption profile can be widened without increasing the doping concentration, e.g., by co-doping with another rare-earth species \cite{Bottger2008}. In this case the number of ions within the reserved frequency range of each qubit decreases, thus reducing the average ISD error that occur per gate operation. This reduction in ISD error can be seen by studying the different concentrations in Fig. \ref{fig:ISD_q_and_op} since decreasing the concentration below roughly $0.5\%$ results in fewer ions per reserved frequency range. For example, the error shown in the red data, which shows between $1$ and $10$ gate operations being applied on only $1$ qubit, decreases as the concentration is lowered from the critical value of around $0.5\%$. When this reduced error per gate is combined with the linear scaling discussed above and the potential saturation of the ISD error due to the limited lifetime of the excited state, it seems possible to reduce the ISD errors compared to the results shown in Fig. \ref{fig:ISD_q_and_op}.


\bibliography{Ref_lib}

\end{document}